\documentclass[12pt,preprint]{aastex}

\usepackage{color}

\newcommand{\msun}{M_\odot} 
 
\def\nuc#1{${}^{#1}$} 
\def\nucm#1#2{{}^{#1}{\rm #2}} 	
\newcommand{\feoh}{[{\rm Fe}/{\rm H}]}
\newcommand{\sufhb}{_{\rm HB}}
\newcommand{\sufenv}{_{\rm env}}

\slugcomment{Revised version of January 31, 2006}
\begin{document}

\title{Various Modes of Helium Mixing in Globular Cluster Giants \\ and Their Possible Effects on the Horizontal Branch Morphology}
\author{Takuma Suda\altaffilmark{1} and Masayuki Y. Fujimoto}
\affil{Department of Physics, Hokkaido University}
\affil{N-10 W-8, Kita-ku, Sapporo 060-0810, Japan}
\email{suda@astro1.sci.hokudai.ac.jp; fujimoto@astro1.sci.hokudai.ac.jp}

\altaffiltext{1}{Meme Media Laboratory, Hokkaido University }

\begin{abstract}
It has been known for a long time that some red giants in
globular clusters exhibit large star-to-star variations in
the abundances of light elements that are not exhibited by
field giants. This fact can be taken as evidence that the
extra mixing mechanism(s) that
operate in globular cluster giants may be consequences of
star-star interactions in the dense stellar environment.
In order to
constrain the extra mixing mechanism(s), we study the
influence of helium enrichment along the red giant
branch (RGB) on the evolution of stars through the
horizontal branch. Three possible modes of helium enrichment
are considered, associated with close
encounters of stars in the globular clusters.
We show that
as a consequence of the variations in the core mass as well
as in the total mass due to mass loss,
the color of horizontal
branch models are distributed over almost all range of
horizontal branch.
The results are discussed in relation to the scenario for
the origin of the abundance anomalies and for the effects on
the morphology of horizontal branch.
We argue that
the star-star interactions can
explain not only the source of angular momentum of rapid
rotation but also provide a mechanism for the bimodal
distribution of rotation rates in some globular clusters.
We also propose the
time elapsed from the latest core collapse phase during the
gravo-thermal oscillations as the second parameter to explain
the variations in HB morphology among the globular clusters.
\end{abstract}


\section{Introduction}
There is growing observational evidence that points to the importance of star-star interactions in globular clusters in modifying stellar evolution, e.g., the smaller relative frequency of red giants in cluster cores, the overabundance of millisecond pulsars, the dependence on the stellar density of populations of blue stragglers \citep[see][for reviews]{Bail95,Hut92}.  
     In particular, the recent inflation of the number of known blue stragglers \citep{Ferr97,Ferr99,Palt01} indicates that a significant fraction of stellar populations may undergo encounters with neighboring stars that result in direct coalescences and/or binary mergers.  
     Additional evidences may be the large star-to-star variations in the abundances of C, N, O, Ne, Mg and Al, which cannot be explained in terms of nucleosynthesis and mixing in the stars within the current standard framework of stellar evolution that takes into account only the thermal convection and molecular diffusion as the mechanisms of element mixing in the interior.  
     Since these anomalous abundances are observed only among giants in globular clusters but not in field halo giants \cite[see][for the review]{Kraf94,DaCo97}, it is natural to search for their origin in the environmental differences between the globular clusters and Galactic halo.  
     It has already been argued that close encounters with nearby stars may responsible for the unusual distributions of red giants and the peculiar morphology of the horizontal branch at least in the central parts of globular clusters  \citep[e.g.,][]{Djor91,Fusi93}.  

One of characteristics of the observed abundance anomalies is that, despite large variations in the abundances of individual elements, the sum of C, N and O abundances and the sum of Mg and Al abundances are found to be constant among giants in each cluster \citep{Smit96,Shet96a,Kraf97}.    
     For C and N and also for O, the abundance variations are found to increase progressively along the RGB \citep{Kraf94,Bell01,Bril02} to an extent which is much greater than observed for field giants \citep[e.g.,][]{Smit03}.  
     These facts suggest that some extra mixing mechanism or mechanisms operate in the course of evolution along the RGB for giants in globular clusters, different for field giants.  
     The origin of the abundance anomalies is further constrained by the discovery of \citet{Shet96b} that, among the magnesium isotopes, \nuc{24}{Mg} is depleted but \nuc{25}Mg and \nuc{26}Mg are not \citep[see also][]{Shet98,Yong03}. 
     Since the burning of \nuc{24}{Mg} demands temperatures much higher than can be reached in the quiescent hydrogen burning shell of red giants \citep{Lang97}, there must be machinery for raising the temperature in the hydrogen-containing layer high enough to burn \nuc{24}{Mg}. 

These observations may constitute the constraints on the nature of the mixing mechanism in cluster red giants, and from proper consideration for them, \citet{Fuji99} propose a scenario for nucleosynthesis and extra mixing which is triggered by a close encounter with nearby stars. 
     It has been shown by SPH simulations that during a close encounter, a significant amount of angular momentum can be transferred into the convective envelope of red giants \citep[][Yamada, Okazaki \& Fujimoto, in preparation]{Davi91}. 
     The deposited angular momentum is transferred inwards effectively by convection to give rise to rotation-induced mixing in the radiative zone below the surface convective zone, and the latter may mix down a small amount of hydrogen from the bottom of hydrogen-burning shell into the upper part of helium core and causes the ignition of hydrogen shell flashes. 
     It is shown that if hydrogen reaches sufficiently deep (over a few pressure scaleheights) into the helium core, the hydrogen-shell flashes can be strong enough to produce the high temperatures necessary to burn \nuc{24}{Mg}. 
     Further, during the decay phase, matter once contained in the convective shell due to the hydrogen flash is dredged up by surface convection, resulting in the enrichment of surface layers with the nuclear products of the hydrogen shell flash. 
     By computing the nucleosynthesis during the hydrogen flashes, it has been demonstrated that this model can reproduce the observed relationship between Na and Al abundances and the anti-correlation between Mg and Al abundances \citep{Aika01,Aika04}.   

Recently, similar abundance variations to those observed from cluster giants have been detected from the turn-off stars and subgiants in NGC 6752 by \citet{Grat01} as the anti-correlations between Na and O and between Mg and Al. 
     The presence of the abundance anomalies in these unevolved stars has been inferred from the variations in the strengths of CN band and from CN and CH anti-correlations extending from giants down to upper main sequence stars (Suntzeff \& Smith 1991 for NGC 6752, Briley et al.~1992 for M5, Cannon et al. 1998 for 47 Tuc, Cohen 1999 for M71), and also from the correlation between CN band strength and Na abundance for 47 Tuc \citep{Bril96}. 
     These facts have tended to be interpreted as refuting the evolutionary scenario that the abundance anomalies are produced during the evolution along the giant branch and as favoring the primordial scenario that the stars were polluted by and/or born of gas already exhibiting the abundance anomalies \citep[e.g., see][]{Sned04}. 
     It is possible, however, to produce the abundance anomalies in unevolved stars through mass transfer from those giants that have already developed the abundance anomalies during the same kinds of star-star interactions that can deposit angular momentum into the convective envelope of giants \citep{Shim03}.   
     Actually, the SPH simulations demonstrate that during a close encounter with a red giant, main sequence stars can accrete envelope mass from the red giant up to $\sim 10^{-2} \msun$ \citep[][Yamada, Okazaki \& Fujimoto, in preparation]{Davi91}, which is enough to disguise their surface convective layer with the accreted material. 

For the metal-poor stars of intermediate masses, it is also shown that the temperature at the bottom of surface convective zone becomes high enough for \nuc{24}{Mg} to burn during the asymptotic giant branch \citep{Vent01,Deni03}.  
     This leads specifically to a recycling scenario of envelope matter ejected from erstwhile AGB stars, which claims that the presently observed low mass stars have been polluted by accreting these ejecta \citep{Thou02} and/or have been born from the polluted gas as second generation stars in the globular clusters \citep{DAnt04,DAnt04b}. 
     Since those stars which undergo hot bottom burning also experience the third dredge-up, however, the pollution is expected to be necessarily attendant with enormous enhancements of CN abundances \citep{Latt04,Fenn04}, and also with large enrichment of s-process elements, as observed in the case of CH stars. 
     This is not the case for CN elements as stated above, and for the s-process elements, the contrary has been observed for the stars in NGC 6752 that exhibit an anti-correlation between Na and O and between Al and Mg \citep{Jame04}. 
     These observations impose difficulties upon the recycling scenario in addition to the obvious one of mass supply; 
     since it takes more than several tenths of solar mass even for giants to disguise their surface with the anomalous abundances, the amount of mass necessary to explain the observations may well exceed the total envelope mass ejected from the erstwhile AGB stars that have experienced hot bottom burning. 

In contrast, the evolutionary scenario may be exempt from these difficulties.  
     During hydrogen shell flashes, neither carbon nor s-process elements are produced. 
     The surface disguise is responsible only for the unevolved stars, and hence, demands the order of mass in their surface convection ($0.001 \sim 0.01 \msun$) since the abundance anomaly can be generated in the stars themselves as soon as the surface convection becomes deep.  
     The evolutionary scenario with star-star interactions taken into account may therefore have a better prospect of success and is worth further investigations. 

In the present work, we assume that deep mixing mechanism(s) operate along the RGB and explore their consequences on the evolution through the horizontal-branch phase. 
   Since, because of a large Coulomb barrier, heavier elements such as Mg and Al can be processed only in the inner part of the hydrogen-burning shell, mixing responsible for Al and Mg abundance anomalies is likely to accompany the concomitant enrichment of surface helium \citep{Lang95}. 
     The effects of possible helium mixing on the evolution have been investigated in relation to the morphology of the horizontal branch \citep{Swei97a,Swei97b,Swei98,Calo01} and to the luminosity function of the cluster giant branch \citep{Lang00,Bono01}. 
     In the existent works, helium mixing has been treated as more or less continuous process. 
     This may not necessarily be true, however, since the extra mixing mechanism has yet to be properly established. 
     In actuality, two different scenarios of helium enrichment may emerge from the models of deep mixing in globular cluster giants, proposed to date. 
     One may assume that meridional currents and/or the turbulence generated by hydrodynamical instabilities of differential rotation carry matter from the top of the hydrogen-burning shell into the surface convective zone \citep{Swei79,Smit92,Char92,Deni00}. 
     Or, one may assume that hydrogen shell flashes are involved, triggered by inward mixing of a small amount of hydrogen into the upper helium core presumably via the turbulence associated with the differential rotation, as proposed by \citet{Fuji99}. 
     In the first scenario, the mixing may occur without interruption and can be approximated as a continuous process. 
     In the second scenario, on the other hand, the mixing has to be intermittent.  
     Although both types of mixing are conceivable in principle, it is only the latter mechanism that has been demonstrated to be able to reproduce the abundance anomalies of Al and Mg and the relationships between Na and Al and between Mg and Al observed in globular cluster giants \citep{Aika01,Aika04}. 
     But it is also true that with the latter mechanism with high temperatures alone cannot explain the observed abundance variations of C and O much larger than that of Mg;  
     there should be supplementary mechanism to mix matter processed under such low temperatures as C and O may burn but \nuc{24}Mg may not \citep{Cava98}, which can be either due to the intermittent mixing with weak flashes of smaller temperatures or to the continuous mixing from the quiescent hydrogen shell burning, or both \citep{Fuji99}.
     The difference in the modes of helium enrichment associated with these extra-mixing mechanisms produce different effects on stellar evolution. 
     From these latter differences we may be able to determine which mechanism of deep mixing nature actually prefers. 

The purpose of our work is to investigate the influences that the different modes of helium mixing may exert on the evolution of red giant and horizontal branches. 
     In addition to the above-mentioned two modes of helium enrichment, we propose an additional mechanism that may operate after the major helium core flash is ignited at the tip of the RGB. 
     These three modes are considered with particular attention to the growth in the core mass and the mass loss along the RGB and to the resultant variations in the luminosity and surface temperature of horizontal branch stars. 
     In this paper, we investigate their general characteristics and provide a new framework for understanding the observed properties.  

In the current framework of the theory of stellar structure and evolution, one of the major factors that have not yet been very well incorporated is the stellar rotation.  
     Since it is generally thought that the extra mixing mechanisms in the stellar interior are associated with rotation, the stars with abundance anomalies are also expected to provide important insights into the role of rotation on internal mixing \citep[e.g., see][]{Tass00,Pins97}. 
     Indeed, for the globular clusters with large star-to-star variations of abundances such as M13, M92, M15, and M79, some horizontal branch stars are observed to rotate rapidly at velocity $v_{\rm rot} \sin i \gtrsim 30 \hbox{ km s}^{-1}$ (where $i$ is the inclination angle of spin axis) \citep{Pete83,Pete95,Cohe97,Behr00a,Behr00b,Reci02}. 
     The origin of angular momentum necessary for such rapid rotation is itself difficult to understand because the angular momentum ought to be effectively extracted during the main sequence phase by magnetic braking \citep{Skum72} and also during the red giant phase by mass loss. 
     In fact, for turn-off stars and subgiants in some globular clusters, an upper bound on the rotation velocity is found as small as $v_{\rm rot}\sin i \lesssim 3.5 \hbox{ and } 4.7 \hbox{ km s}^{-1}$ \citep{Luca03}. 
     The problem of the source of angular momentum has been addressed in relation to the storage of angular momentum in the stellar core by \citet{Sill00} and in relation to spin-up as a result of swallowing close planetary companion during the RGB evolution by \citet{Soke98}. 
     As criticized by \citet{Reci02}, however, these scenarios suffer from defects; 
     the first cannot specify an adequate source of large angular momentum in the core, and the second stumbles over the failure to detect the very existence of planets in an intense search in 47 Tuc by \citet{Gill00}.  
     Furthermore, horizontal branch stars are known to display a bimodal distribution of rotation velocity, in which rapid rotation is restricted to the cooler horizontal branch stars ($T_{\rm eff} < 15000^\circ$K), whereas the slower rotators ($\lesssim 10 \hbox{ km s}^{-1}$) spread over a wider range of effective temperature.   
     This is at variance with the general expectation that faster rotation may lead to larger core and loss of larger envelope mass during RGB to shift the star bluer wards on HB.   
     Consequently, there should be mechanism(s) not only to supply the angular momentum but also to bring about this bimodal distribution.  
     In our scenario, the star-star interaction in a dense stellar environment is expected to be a possible source of angular momentum in rapidly rotating HB stars.
     In addition, we show that our scenario can provide an explanation also for the bimodal distribution of angular velocity in horizontal branch stars.     

The organization of the paper is as follows:  
      In \S 2, we elaborate the modes of helium mixing to be considered and broadly classify them into three typical modes.       
     In \S 3, we summarize the method and assumptions of the computations.   
     The consequences of numerical simulations are presented in \S 4.  
     Conclusions follow in \S 5 with a discussion, though preliminary, on the relevance of our results to some of the observed properties of globular clusters. 
    Detailed comparisons with the observations demand the specification of various parameters from cluster to cluster and will be discussed in a subsequent paper.

\section{Three Modes of Mixing for Surface Enrichment of Helium}
We consider three different modes of helium mixing in this paper.  
    The first two are associated with the extra mixing mechanisms, suggested by other authors to work in cluster red giants, and differ primarily with regard to the time variations;    
    one may regarded as continuous process that occurs without interrupting hydrogen shell burning and the other as intermittent process that recurs with hydrogen shell flashes. 
    The last one is proposed to operate after the helium core flash ignites at the tip of RGB when the convective zone, driven by helium burning, extends outward into hydrogen-containing layers and engulfs hydrogen.  
     We start with formulating the characteristics of three modes of helium enrichment discussed in this work. 

\subsection{The Continuous Mixing Mode} 
Continuous mixing applies to most of the extant deep mixing models which deal with mixing outward from the quiescent hydrogen burning shell, as studied e.g., by \citet{Swei79}.  
     Mixing processes of this type may be properly described in terms of the strength of the mixing current and its depth of penetration into the hydrogen-burning shell \citep{Boot95,Weis00,Deni00}. 
     Because of the lack of a proper theory to constrain them, these two quantities must presently be deemed free parameters. 
     Since nevertheless, the rate of helium mixing into the envelope is relevant to our problem, we adopt a simpler model with only one parameter that specifies the amount of helium produced in the hydrogen-burning shell that is carried out into the envelope.  
     Denoting this parameter as $f$, we write the rate of change in the hydrogen abundance, $X$, in a shell as 
\begin{equation}
\left( {\partial{X} \over \partial{t}} \right) _{M _r} = - \frac{\varepsilon_{\rm H}} {E_{\rm H}}(1-f), 
\label{eqX} 
\end{equation}  
     and the rate of change of the hydrogen and helium abundances, $X_{\rm env}$ and $Y_{\rm env}$, in the envelope as 
\begin{equation}
 \frac{ d X_{\rm env}}{d t} = - \frac {d Y_{\rm env}} {d t } =  - \frac{f L_{\rm H}} {(M - M_{1e}) E_{\rm H}}.   
\label{eqY}
\end{equation}  
    Here $\varepsilon_{\rm H}$, $L_{\rm H}$ and $E_{\rm H}$ are, respectively, the nuclear energy-generation rate per unit mass, the total energy-generation rate in the star due to hydrogen burning, and the energy release per unit mass of hydrogen. 
     The outer edge of hydrogen-burning shell $M_{1e}$ is defined as the point where $X = X_{\rm env} - 0.01$.  

It should be noted that mixing due to the meridional circulation is inhibited by a gradient in the mean molecular weight \citep{Mest57}.  
     From the energetics, the condition for mixing across layers which differ in the mean molecular weight by $D \mu$ is 
\begin{equation}
\chi _r \equiv \left (\Omega \over \Omega_{\rm K} \right)^2 > \left( {2 H_P \over r } \right) {\vert D \mu \vert \over \mu} 
\label{mmbar}
\end{equation}
\citep{Kipp74}, where $\Omega$ is rotation rate at the shell, $\Omega_{\rm K} \ (= [ GM_r / r^3]^{1/2})$ is the local critical rotation rate, and $H_P \ (= - d r / d \log P)$ is the pressure scale height.   
     A similar condition is applicable to mixing by hydrodynamical instabilities due to differential rotation, with the rotation rate replaced by the differential rotation rate \citep{Fuji88}.  
     In the envelope of red giants, a discontinuity in the abundance profile is formed at the first dredge-up. 
     This discontinuity can be as large as $\Delta X \simeq 0.05$ with $H_p \ge r / 4$.  
     It has been argued, therefore, that the mixing of nuclear products from the hydrogen-burning shell to the surface can occur only after the hydrogen-burning shell passes through the discontinuity unless the shell rotates at rates close to the local critical rotation rate and $\chi > 0.02$.  
     This condition may also be applied to the mixing from the hydrogen-burning shell, across which there is a steep variation in mean molecular weight.  
     Accordingly, a mixing current in the upper radiative zone penetrates into the hydrogen-burning shell only by a small amount, and hence, it is unlikely to mix an appreciable amount of helium into the envelope or to cause large variations in heavier elements such as Mg, and Al except for very rapid rotation close to the local critical rate, as discussed by \citet{Swei79}.  

\subsection{The Intermittent Mixing Mode} 
The flash-assisted deep mixing model, proposed by \citet{Fuji99}, also postulates some extra mixing mechanism as in the other extant mixing models but it works in the opposite direction to the others. 
     Instead of the direct outward transport of nuclear products from the hydrogen-burning shell, this model focuses on the inward transport of hydrogen from the bottom tail of the hydrogen-burning shell into the upper helium core and deals with the events consequent upon this mixing. 
     In red giants, matter undergoes a large radial shrinkage while it leaves the bottom of the surface convective zone, passes through the hydrogen-burning shell and finally becomes incorporated in the helium core.  
     If the matter has sufficient angular momentum, therefore, differential rotation ought to be enhanced to trigger hydrodynamical instabilities.  
     The turbulence generated by these instabilities will prevail not only in layers exterior to the hydrogen burning shell but also in layers in interior region below the shell; these interior layers are separated from the envelope only by a very thin (much less than a pressure scale height) layer, and it is expected that the turbulence will cause material mixing between the tail of the hydrogen-burning shell and the top of helium core. 
     If the turbulence is sufficiently strong, a small amount ($X \simeq 0.001$) of hydrogen can be carried down into the helium core sufficiently deep (by a few pressure scale heights) in a sufficiently short time scale ($10^2 \sim 10^3$ yr) that eventually hydrogen shell flash will be ignited.  
     Because of a small abundance of hydrogen involved and also of small ratio of the pressure scale height to the radial distance ($H_P / r \lesssim 0.1$) below the hydrogen burning shell, the barrier of mean molecular weight as discussed in eq.~(\ref{mmbar}) is much abated in comparison with the outward mixing from the top of hydrogen burning shell.  

The mixing process associated with the shell flash may be outlined as follows:    
     The flash, once ignited, drives convection outward into the upper hydrogen-containing layers and strengthens as newly engulfed hydrogen as fuel, to extend it even beyond the site of the now extinct, quiescent hydrogen-burning shell. 
     Along with the addition of entropy, the burning shell expands to reduce the pressure therein, and hence, the flash climbs over the peak stage and starts to decrease the temperature. 
     As the hydrogen burning rate decreases, the flash convection retreats, and finally, the model star settles into a quiescent phase of stable hydrogen shell-burning again. 
     Since the hydrogen burning shell shifts inwards during these events, the base of convective envelope moves inward in mass and may penetrate into the region that has been occupied by the flash convection to dredge up freshly synthesized nuclei, helium as well as the nuclear products during the quiescent hydrogen shell burning and during the flash under high temperatures. 

At the same time, the distribution of rotational velocity is modified both through the efficient transport of angular momentum by convection and through the expansion of the burning shell. 
    The mixing of hydrogen is consequent upon the inward transfer of angular momentum via turbulent diffusion.  
    During the following shell flash, the differential rotation will be largely eliminated in the shells involved in the flash-driven convective zone;  
    the angular momentum is further carried inward, though partly carried back outward due to the expansion of the burning shell.  
     After the dredge-up by the surface convection, a sharp discontinuity in the rotational velocity forms at the interface between the base of the convective envelope and the outer edge of the erstwhile flash-driven convective zone, left behind the dredge-up (unless the top of the helium core and the bottom of convective envelope have similar angular momentum). 
   There generated is strong differential rotation, and hence, violent turbulence; the latter will trigger the inward mixing of hydrogen again to ignite another hydrogen shell flash as the hydrogen burning shell traverses the site of the former flash-driven convective zone and approaches to this interface. 
     In this way, the events of flash-assisted deep mixing may recurs, and this epoch of recurrent flashes and mixings is expected to last until the initial difference in specific angular momentum between the top of the helium core and the bottom of the envelope convective zone is eliminated. 
    Although the mass of the flash-driven convective zone may be much smaller than that of the envelope, a number of shell flashes during the mixing epoch may be able to produce observed variations in surface abundances if the initial difference in the angular momentum is sufficiently large,

In conclusion, for the flash-assisted mixing model, the helium mixing occurs intermittently in a short time at the end of shell flashes unlike the other extant models of deep mixing. 
During the mixing epoch, the helium core may suffer from complicated
changes in the period of alternative hydrogen flashes and quiescent burnings.  
At the ignition of a shell flash, the hydrogen-burning shell
shifts inward and the core mass decreases abruptly,
while the core mass changes very little after the ignition
because of the short time-scale to burn hydrogen.
During the quiescent burning after the flash, the helium
core grows at the same rate that hydrogen burns
in the shell until the next shell flash is ignited.   
     The latter stable burning phase persists much longer than the flash, as estimated from the expression
\begin{equation}
t_{\rm rec} \sim 10^4 (\Delta M_{\rm f c, r} / 0.001 \msun) (X_{\rm f c, r} / 0.1)/(L_H / 10^3 L_\odot) \hbox{ yr}. 
\end{equation} 
Here $\Delta M_{\rm f c, r}$ and $X_{\rm f c, r}$ are the
mass and hydrogen abundance of the hydrogen flash-driven
convective zone, left behind the dredge-up, which may vary
with the efficiency of inward hydrogen mixing and the
strength of hydrogen shell flashes, i.e., according to
how deep the hydrogen carried in, how far the flash-driven
convection erodes the hydrogen-rich layers and how much
mass is dredged up by the envelope convection.  
     The resultant change in the core mass through the recurrent episodes in the mixing epoch depends on these factors, and it may happen that the core mass even decreases if the flashes are sufficiently strong so that the convective envelope penetrates into the site of the flash-driven convection beyond the erstwhile hydrogen-burning shell during the preceding stable quiescent phase. 
     Precise modeling of these processes may, however, be beyond our current understandings on the hydrodynamical instabilities and the turbulence in the rotating stars. 

The thermal state in the core may also be affected by the behavior of the core during the mixing epoch. 
     When a hydrogen shell flash ignites, the core is deprived of the upper shell of higher entropy and suffers from cooling at first, and, then, is heated as heat flows inward from the burning shell as a temperature inversion develops due to the flash. 
     During the stable burning phase between flashes, the core undergoes compression at the same rate as hydrogen burns in the shell, which is larger as compared with standard evolution because of smaller hydrogen abundance in the remnant site of flash convection zone than in the envelope. 
     The competition between these processes leads to a thermal balance in the core during the mixing epoch.  
    It is worth noting that a similar situation occurs in thermally pulsating AGB wherein recurrent helium shell flashes interrupt hydrogen shell burning and dredge-up decreases the core mass. 
     In the AGB case, it has been shown that the thermal state in the core is little affected by the intervening helium shell flashes and is determined primarily by the compression due to hydrogen shell burning \citep{Fuji79}.  

Since the necessary equipment is an electron degenerate core and a deep convective envelope, flash-assisted deep mixing can work as soon as the stars evolve into red giants and even before the hydrogen burning shell reaches the discontinuity of hydrogen profile produced by the first dredge-up. 
     An additional requirement is a strong gradient in the rotation velocity to generate turbulence and to induce the inward mixing of hydrogen, which may in turn demand a large amount of angular momentum in the envelope. 
     We may well attribute the latter to the deposition of angular momentum into the envelope of red giant through the tidal interactions at close encounters with neighboring stars in the dense environment of globular clusters \citep{Fuji99}.  
     This interpretation is in accordance with the fact that large abundance anomalies have been observed for cluster giants but not for field giants in the Galactic halo \citep{Kraf97}.   
     This scenario also implies that the epoch of helium mixing may start at any stage of RGB evolution, being triggered by a close encounter.  

\subsection {The Helium Flash-Driven Deep Mixing Mode} 
When the helium core flash is ignited at the tip of the RGB,
convection is driven outward by the helium burning
and approaches the tail of the hydrogen-containing layer.  
     For extremely metal-poor stars ($[{\rm Fe}/{\rm H}] < - 4$), it is known that helium flash-driven convection extends through the bottom of hydrogen rich envelope and engulfs hydrogen (Fujimoto et al.\ 1990, 1995, see also Fujimoto et al.\ 2000). 
     For Population I and II stars, however, the outer edge fails to contact the hydrogen-rich envelope, but only by a few pressure scale heights, within the current standard framework of stellar evolution.  
     For a $M = 0.8 M_\odot$ star with $[{\rm Fe}/{\rm H}] = -1.5$, the outer edge of the helium convective zone comes nearest, by 1.4 times a pressure scale-height, to the bottom of the hydrogen-rich envelope when the helium-burning rate decreases to $L_{\rm He} = 8.8 \times 10^4 L_\odot$ at 29.6 yr after the peak of helium burning rate of $L_{\rm He} = 4.1 \times 10^9 L_\odot$,. 
     This shortest distance between the outer edge of the helium flash-driven convective zone and the tail of hydrogen containing layer is $\Delta M = 2.1 \times 10^{-3} \msun$ in mass, which corresponds to $\sim 3$ pressure scale-heights during the stationary burning phase just before the ignition of helium flash. 

On the other hand, we have argued in the preceding subsection for a mechanism that may bring about the inward mixing of hydrogen to trigger the flash-assisted deep mixing. 
     Indeed, \citet{Aika01} have concluded that the inward mixing of hydrogen over a few ($2.5 \sim 3.5$) pressure scale-heights is necessary to explain the abundance anomalies observed from M13 giants.  
    If the same extra mixing mechanism works at the tip of the RGB, helium flash-driven convection can reach the hydrogen-containing layers to carry the matter inward.
     It is true that angular momentum is thought to be removed effectively from stars through mass loss along the RGB. 
     And yet, the stars that experience a close encounter with another star at late stage of RGB evolution can retain sufficient angular momentum even at the tip of the RGB. 
     If this is the case, we may expect the hydrogen injection into the helium flash-driven convective zone even for stars of younger populations.  

Hydrogen, thus mixed into the helium flash-driven convective zone, is further carried down by convection to burn and ignite a shell flash in the middle of the zone, with consequences similar to those demonstrated by \citet{Holl90} for a population III star. 
     Once ignited, hydrogen shell flash develops to splits the convective zone into two zones, the upper one driven by the hydrogen shell flash and the lower one by the helium shell flash. 
     During the decay phase of the hydrogen shell flash, the flash convection retreats to disappear and as the burning shell expands, the surface convection deepens in mass to penetrate into the region where flash driven convection prevailed earlier. 
     Although this sequence of events occurs only once, and yet, the decrease in core mass can be large enough to bring about appreciable changes in the surface abundance, as shown later. 
     We call this dredge-up process, induced by hydrogen mixing into the helium-flash convection, as helium flash-driven deep mixing (abbreviated as He-FDDM in the following).  

\section{Numerical Method and Approximations} 
We compute the evolution of low mass stars along the RGB and the horizontal branch, taking into account the possible changes in the surface helium abundance according to the three modes of mixing discussed above. 
   For the helium mixing processes, we adopt simplifying prescriptions to make them tractable by a one-dimensional stellar code. 
   The initial stellar mass is set at $0.8 \msun$ and the initial helium abundance at $Y=0.24$. 
   The metallicity is chosen to be $[{\rm Fe}/{\rm H}] = -1.5$ with an 0.4 dex enhancement of $\alpha$ elements ($Z = 6.8 \times 10^{-4}$), relevant to the moderately metal-poor globular cluster M13, which has been extensively observed and is known to show large star-to-star variations in surface abundances. 
   The initial CNO abundances are taken to be $X_{12} = 7.0 \times 10^{-5}$, $X_{14} = 2.7 \times 10^{-5}$, and $X_{16} = 5.7 \times 10^{-4}$.  
   Other input physics are the same as in \citet{Iben92} since we are interested in how the properties of the mixing models depend on the core mass at the time of the helium core flash.  

In the case of continuous mixing, we compute the variations in the abundances from eqs.~(\ref{eqX}) and (\ref{eqY}) with the parameter $f$ set constant. 
   The mixing is started when the surface luminosity has reached $\log L / L_\odot = 2.1$ (corresponding to core mass of $M_1 = 0.3058 \msun$) and is continued for the rest of the RGB life until the helium-burning rate reaches $1 L_\odot$. 
   This choice is for the sake of simplicity and for comparison with other computations \citep[e.g.,][]{Swei97a}. 
   In our model star, the discontinuity of abundance in the envelope formed by the first dredge-up is located at $M_r = 0.3252 \msun$. 

In the case of flash-assisted deep mixing, it is of little use to follow the detailed process of hydrogen mixing since we currently lack the knowledge necessary to model rotation-induced mixing definitively. 
   For the same reason, we ignore the change in core mass during the mixing epoch as well as the influence on the thermal state of the core, as discussed in the preceding section. 
   Instead, we simply increase the helium abundance to a given value in a short timescale with the thermal state of the core little changed. 
   Flash-assisted deep mixing may start at any time on the RGB when the giant undergoes a close encounter with another star, as postulated in the scenario by \citet{Fuji99}. 
   In order to see the dependence on the evolutionary stage when mixing occurs, we compute two cases with mixing occurrence at two different epochs: (1) the same stage ($\log L / L_\odot = 2.1$) as the continuous mixing model starts and (2) a later stage ($\log L / L_\odot = 2.6$, when $M_1 = 0.3704 \msun$). 
   The time span between these two stages is 21.4 Myr, about two thirds of the RGB lifetime from the stage of luminosity $\log L / L_\odot = 2.1$ to the tip of RGB in the no-mixing model. 

In the case of He-FDDM, again for the same reason, we do not follow the process of hydrogen engulfment. 
   Instead, we assume simply that a sufficient amount of hydrogen is mixed and carried down in the helium convection at the stage of the maximum outward extension of helium convection to ignite hydrogen shell flash. 
   The location of the shell where the hydrogen flash ignites is estimated by equating the lifetime of protons to the local convective turnover timescale. 
   For obtaining the proton lifetime in the helium convective zone (at the stage of maximum helium convection), relevant reactions are $\nucm{12}{C}(p, \gamma) \nucm{13}{N}$ and/or $\nucm{18}{O}(p, \alpha) \nucm{15}{N}$, where the abundances by mass of \nuc{12}C and \nuc{18}O are set at 0.041 and 0.00068, respectively. 
   Further, for simplicity, we assign a uniform distribution of hydrogen in the helium convective zone above the shell that hydrogen-flash ignites, and to mimic the process of hydrogen engulfment, increase the abundance therein from $X = 10^{-7}$ to given abundances of $X_{\rm mix}= 0.001$ and 0.01 in $\sim 10^7$ s. 
     Then we follow the development of hydrogen shell flash and subsequent dredge-up by the envelope convection within the standard framework of stellar evolution.  

\section{ Results of Evolutionary Computations }
Table~1 summarizes the model parameters with the resultant surface abundance, $Y_{\rm env}$, of helium, the core mass, $M_{1, \rm ZH}$, when they settle on the horizontal branch, the peak luminosity, $L_{\rm tip}$, at the tip of RGB, the mass coordinate, $M_{\rm 2,i}$, at the bottom of helium burning shell during the major helium flash, and the amount of mass loss, $\Delta M_{\rm ml}$, through the RGB evolution. 
   Here and in the following, the core is defined as the interior to the mass shell where the hydrogen abundance reduces to a half of the surface abundance, i.e., $X = X_{\rm env}/2$. 
   These models differ only after the onset of helium mixing on the RGB. 
   When they reach the base of the RGB, the surface helium abundance has already increased owing to the first dredge-up, as seen from the model without helium mixing, the difference from $Y_{\rm env} = 0.2491$ gives the degree of surface helium enrichment due to the deep mixing.  

In the following subsections, we first discuss the characteristics of RGB evolution for the each mixing model, and then, examine their effects on the morphology of horizontal branch with the possible influence of mass loss taken into account.  

\subsection{Characteristics through Red Giant Branch evolution }
\subsubsection{The Continuous Mixing Models}
In the continuous mixing models, the helium mixing influences the growth rate of core in two opposite ways, i.e., it cut downs the helium addition to the core below the hydrogen burning shell while it enhances the hydrogen burning rate as a result of the decrease in the hydrogen abundance in the envelope.  
   Figure~1 shows the time variations in the surface luminosity for various values of helium mixing rate, $f$.  
   When the helium mixing starts, the increase of luminosity slows down to greater extent for larger rate of helium mixing $f$, which is a direct consequence of the delay of core growth by a factor of $1-f$.  
   Then, as the helium abundance in the envelope augments by the helium mixing, the hydrogen burning rate increases, and finally, becomes larger than that of the no-mixing model at the tip of RGB.  

Figure~2 shows the variations in the surface helium abundance, the hydrogen burning rates, and the growth rates of core mass as a function of the core mass.  
   Since we assume constant mixing rate $f$, the helium abundance, $Y_{\rm env}$, in the envelope increases almost linearly with the core mass as seen from eqs.~(\ref{eqX}) and (\ref{eqY}) ($d \log [1-Y_{\rm env}-Z]  = f/(1-f) d \log [M-M_{1 e}]$ with the effect of mass loss neglect).  
    When compared at the same core masses, the hydrogen-burning rate is larger for larger helium mixing rate since the opacity decreases with the hydrogen abundance in the lower envelope adjacent to the hydrogen burning shell. 
     Nevertheless, during the early stages, the growth rate of core mass remains smaller for the models of larger mixing rate because of larger deduction in the helium addition to the core. 
     As the surface helium abundance increases, the growth rate of core rises owing both to the enhancement of hydrogen-burning rate and to the reduction in the nuclear energy release from unit envelope mass, eventually to exceed that of no-mixing model at $ M_1 \simeq 0.44 \sim 0.40 \msun$ (see the analytical solution of eqs.~{\ref{eqX}} and {\ref{eqY}} in Appendix A).  
Since the thermal state in the core is determined by the growth rate of core, the evolution to the ignition of helium core is consequent upon the competition between these two effects. 

Figure~3 illustrates the thermal state in the core on the density and temperature diagram for different mixing rates.  
   For the models of larger mixing rates, the temperature in the hydrogen burning shell is higher so as to compensate the reduction in the pressure of hydrogen burning shell with decreasing hydrogen abundance in the envelope, which results in an increase in the hydrogen burning rate to meet the concomitant decrease in the opacity near the burning shell. 
   In contrast, the central temperature and entropy become lower in the models with larger mixing rate after the mixing is switched on, for the thermal state in the inner part of core ensues from the balance between the compression due to the increase in the core mass and the heat loss from the core.  
   After the core mass grows more massive than $M_1 \simeq 0.30 \msun$, the temperature inversion develops because of neutrino losses from the center, and the maximum temperature comes to occur in the middle of core, near the shell with $\partial \log \kappa / \partial P \mid _T = -1$, which is defined as watershed by \citet{Fuji84} since neither the radiation heat transfer nor the heat conduction by electrons is least efficient there. 
   In the models of larger mixing rate, the maximum temperature stays smaller even after the growth rate of core resumes to be greater than that in the no-mixing model, as seen from the models with $M_1 = 0.4 \msun$;  
it takes a quite while before the effects of preceding
cooling in the central part are wiped away by the
compression of core at the enhanced rate, for the inward
heat flow from the watershed depends sensitively on the
temperature and the heating of central part is not so efficient process.  

Consequently, the ignition of helium core flash is delayed until a larger core mass is reached for larger mixing rate in the range of $f \le 0.4$, and for still larger mixing rate, the core mass at the helium ignition turns to decrease, as seen from the model of $f = 0.6$.  
Because of the cancellation between the above mentioned two effects, however, the difference in the core mass is rather small, as pointed out by \citet{Swei97a};  
in our case. $\Delta M_{1, \rm ZH} \le 0.0059 M_\odot$ and $\Delta M_{1, \rm ZH}/\Delta Y{\rm env} \le 0.0378$.  
     For larger mixing rates, the helium flash ignites in the outer shell because the watershed shifts outer in the mass fraction as the central density increases with the core mass and so does the shell of maximum temperature;    
     the helium burning shell, or the bottom of helium convective zone, during the major core flash varies from $M_{2i}/ M_{\rm 1, ZH} = 0.447$ in the no-mixing model to $0.454$ and $0.539$ for the models of $f=0.20$ and $f=0.6$, respectively. 

\subsubsection{The Intermittent Mixing Models} 
The intermittent mixing differs from continuous mixing in that the core mass grows at the same rate as the hydrogen burning in the shell except for the short intermission due to the hydrogen shell flashes.  
   Figures 4 and 5 show the time variations in the luminosity and the hydrogen-burning rate as a function of the core mass for models of intermittent mixing, respectively. 
   The models denoted by solid and broken lines are constructed under the assumption that the thermal state in the core little changes during the mixing epoch with the latter duration artificially shortened, as discussed above  
   After the mixing epoch, the hydrogen-burning rate, which has increased with the helium enrichment, raises the growth rate of core and heats up the core above the no-mixing model. 
   Figure~6 illustrates the evolutionary trajectories of the central part on the density and temperature diagram. 
   Because of the enhanced compression rate of core after the mixing epoch, the central temperature undergoes steeper rise with the central density for larger mixing rate; the whole core is kept hotter from the center through the hydrogen burning shell, and the helium core flash ignites at a smaller core and at an inner shell when compared among the models of the same mixing epoch. 
When compared among models with the same mixing rate,
the heating effect of core is greater for an earlier
occurrence of the mixing epoch, leading to the ignition
of helium core flash at a smaller core mass. 

The tendency of the core mass at the helium ignition decreasing with the helium enrichment is in accordance with the former studies on the effects of different pristine helium abundances \citep[e.g.,][and references therein]{Cate96}.
The core mass $M_{1, \rm ZH}$ at the helium ignition depends
on the surface helium enrichment as
$\Delta M_{1, \rm ZH}/\Delta Y_{\rm env} = -0.19 \sim -0.15 \msun$
for the models of mixing epoch at $\log (L / L_\odot) = 2.1$,
which may be compared with the dependence on the pristine
helium abundance of $d M_{1, \rm ZH } / d Y = -0.24 \msun$
for the no-mixing models \citep{Cate96}.
For a later occurrence of mixing epoch at $\log (L / L_\odot) = 2.6$,
the variation of core mass decreases to
$\Delta M_{1, \rm ZH }/\Delta Y_{\rm env} = -0.11 \msun$. 
   The models of the earlier mixing epoch have smaller luminosities at the tip of RGB than the model without helium mixing despite the surface helium enrichment. 

In the above computations, we ignore the time necessary to produce the amount of helium mixed into the surface for the consistency of the thermal state in the core.  
   In order to assess the elongation of RGB lifetime, we compute the other extreme cases in the same way as in the continuous models but with $f = 1$ until the helium abundance in the envelope reaches to given values.  
     This implies that the mixed helium is solely produced via the quiescent burning at a fixed core mass with concomitant increase in the helium abundance of envelope taken into account, and that the effects of the cooling during the mixing epoch is maximally taken into account.  
   As shown in Figure~4 by dotted lines, it takes 28.3 and 6.8 Myr to produce the amount of helium for the surface enrichment of $\Delta Y_{\rm env} = 0.0793$ with the mixing epoch at $\log (L / L_\odot) = 2.1$ and 2.6, respectively. 
   Consequently, the intermittent mixing models spend the RGB lifetime longer than the no-mixing model by 19.8 and 4.1 Myr, respectively, and the elongation amounts to fairly large fraction of RGB lifetime in particular when the mixing epoch occurs in an early phase of RGB evolution.
   As compared with the continuous mixing model of the same helium enrichment (for which the elongation is 4.4 Myr), the RGB lifetime is much longer when the mixing epoch occurs earlier while it becomes shorter for later occurrence of mixing epoch. 
   For the other models in Fig~4, we add the time necessary to produce the amount of helium mixed into the surface, estimated from the quiescent burning rate at a constant core mass. 
   In these models, the reduction of core mass is alleviated to be much smaller than obtained above ($\Delta M_{1, \rm ZH}/\Delta Y_{\rm env} = - 0.11$ and $-0.02 \msun$ for the mixing epoch at $\log L / L_\odot = 2.1$ and 2.6, respectively).  
   It should be noticed, however, that the effects of the cooing during the mixing epoch is overestimated, as stated in the preceding section. 

\subsubsection{The Helium Flash-Driven Deep Mixing Models}
In the present study, we postulate that an amount of hydrogen is mixed into the helium convective zone sufficiently to split the convection.  
   Figure~7 shows the competition between the lifetime of protons and the convective turnover time scale in the helium convective zone as a function of mass coordinate for the stage of maximum extension of helium convection. 
     Here the reaction rates are taken from \citet{Caug88} and the convective turnover timescale is evaluated from the mixing length theory with the same mixing length in the evolution computations.  
   The lifetime of protons decreases steeply with the depth while the convective turnover timescale varies slowly, and they meet to be comparable at the shell of $M_r \simeq 0.30 \msun$ at the stage of maximum extension of helium convection.  
   Assuming that the engulfed hydrogen is carried down to burn in the shell of $M_r = 0.30 M_\odot$, we compute two evolutionary sequence with artificially increasing the hydrogen abundance in the hydrogen-flash convection to $X_{\rm mix} = 0.001$ and 0.01. 

Figure~8 shows the variations in the luminosity and in the hydrogen-burning rate against the core mass, $M_1$; the core is defined as interior to the shell where the hydrogen abundance is a half of the surface value, as stated above, and hence, the core mass varies with the extension of convection as well as the shell burning.   
   When the hydrogen abundance is increased to $X = 6.4 \times 10^{-7}$ in the mixed shells, the hydrogen burning rate increases to $L_{\rm H} = 6.4 \times 10^6 L_\odot$ to split the convection into two zones. 
   The upper one, driven by the hydrogen shell flash, develops further outward, while the lower one, driven by the helium shell flash, soon disappears owing to the decrease in the weight of overlying layer due to the hydrogen shell flash (when $L_H = 2.9 \times 10^7 L_\odot$). 
   For the model of $X_{\rm mix} = 0.001$, the hydrogen shell flash reaches the peak of $L_H ^{\rm max} = 3.1 \times 10^8 L_\odot$. 
   Even after this stage, the hydrogen convection continues to extend outward, and eventually reaches to the shell of $M_r = 0.49067 \msun$ and $X = 0.748$ beyond the former, now extinct, hydrogen burning shell (of $M_r = 0.4903 \msun$ when $L_H = 9.5 \times 10^4 L_\odot$), as seen from this figure. 
   The erosion of overlying hydrogen-rich layer raises the hydrogen abundance in the convective zone to $X= 0.0016$. 
   Along with the subsequent expansion of hydrogen burning shell, the flash convection retreats finally to disappear and the hydrogen-shell burning settles in a stable quasi-equilibrium state at $\log L_{\rm H} / L_{\sun} \simeq 2.5$. 
   At the same time, the surface convection deepens in mass to penetrates into the former site of hydrogen flash-driven convection down to $M_r = 0.4745 \msun$ after $2.6 \times 10^3$ yrs from the onset of hydrogen mixing, which enriches the surface helium $\Delta Y_{\rm env} = 0.0354$. 
   This deepening of surface convection is consequent upon the inward shift of hydrogen-burning shell due to the hydrogen mixing and the shell flash (into the shell of $M_r \simeq 0.30 \msun$). 
    In this model, however, the hydrogen burning shell passes over the site of hydrogen-flash convection, left behind by the dredge-up, in relatively short time ($\sim 3 \times 10^5$ yr) because of very small hydrogen abundance. 
   Accordingly, the model thus enters into the horizontal branch with the core mass, $M_{1, \rm ZH} = 0.4745 \msun$. 

For the model of $X_{\rm mix} = 0.01$, the hydrogen shell flash is slightly stronger ($L_H ^{\rm max} = 4.7 \times 10^8 L_\odot$), and drives the flash convection more outward up to the shell of $0.4911 M_\odot$ (when $L_H = 9.5 \times 10^4 L_\odot$), as seen in Fig.~8 with the hydrogen abundance in the flash convection amounting to $X = 0.0123$. 
   After the flash, the hydrogen shell burning settles in the stable quiescent state at the very bottom of the site of hydrogen flash convection since the nuclear timescale ($\tau_{\rm nuc} = X / \varepsilon _H \simeq 10^{12}$ s) is much longer than the heat diffusion timescale ($\tau_{\rm dif} \simeq 4 \times 10^{10}$ s), differently from the model of $X_{\rm mix} = 0.001$. 
   Accordingly, the surface convection reaches deeper in mass down to the shell of $M_r = 0.4211 M_\odot$, which enriches the surface helium $\Delta Y_{\rm env} = 0.1313$. 

The distinguishing feature of this mode of mixing is the concomitant enrichment of carbon and nitrogen, produced by the helium flash and processed by the hydrogen flash. 
   Since the carbon abundance is $X_{\rm C} = 0.041$ at the onset of hydrogen mixing, the CN enrichment in proportion to the helium enrichment as $\Delta X_{\rm C N} / \Delta Y_{\rm env} \simeq 0.041 / (0.96 - 0.25) = 0.058$. 
   In our models, the carbon and nitrogen abundance in the envelope increase by factors of 26 and 19 to $X_{12} = 1.73 \times 10^{-3}$ and $X_{14} = 5.0 \times 10^{-4}$ for the model of $X_{\rm mix} = 0.001$, and up to $X_{12}= 6.2 \times 10^{-3}$, $X_{14}= 1.845 \times 10^{-3}$ for the model of $X_{\rm mix} = 0.01$.  
   Because of dredge-up, the oxygen abundance decreases slightly to $X_{16} = 5.4 \times 10^{-4}$ and $X_{16}= 4.65 \times 10^{-4}$, respectively.  

If the hydrogen mixing occurs at an earlier stage before the helium convection reaches its maximum extension, the mixed protons will burn in upper mass shells, as seen from Fig.~7; 
   during the decaying phase of helium flash, both the temperature and the helium burning rate in the helium convective zone decline with time, and yet, since the temperature decrease is larger in the middle than in the bottom of convective zone, the increase in the lifetime of proton capture is larger than that of the convective turnover timescale. 
   In the case of earlier occurrence, however, the pressure is also higher in the shell where protons burn so that the hydrogen flash is stronger to expel the hydrogen flash convection more outward. 
   This entails larger hydrogen abundance in the hydrogen-flash convective zone, which in turn promotes the penetration of envelope convection when the hydrogen-burning shell settles in a stable burning state. 
   Because of this compensation, the resultant core mass will not differ so much when the model reaches on the horizontal branch. 

\subsection{Characteristics through Horizontal Branch Evolution}   
The relationship between the helium enrichment and the core mass when the model stars reach the horizontal branch, is summarized in Figure~9 for the three different modes of helium mixing. 
   For the continuous mixing models, the core mass increases with the helium mixing rate for $f \lesssim 0.4$, and levels off or decreases for still greater mixing rate.  
   The continuous mixing ends with the cores more massive than the no-mixing model and the largest ones among the three modes of helium mixing. 
   In contrast, the intermittent mixing yields a smaller core than the no-mixing model; the core mass is a decrease function of the helium enrichment, and smaller for an earlier occurrence of mixing epoch. 
   The differences in the core mass from the no-mixing model are much larger than for the continuous mixing models with the same helium enrichment. 
The He-FDDM gives the smallest core masses among the three modes of helium mixing and sets the lower limit to the core mass for a give helium enrichment since there is no additional production of helium to make up for the fraction carried out into the envelope, i.e., 
\begin{equation}
\Delta M_{1, \rm ZH} / \Delta Y_{\rm env} \ge - M_{\rm env, n m}/(0.96 - Y_{\rm env} - \Delta Y_{\rm env}) \simeq -0.45 \sim -0.53 
\end{equation} 
  in our cases, where $M_{\rm env, n m}$ is the envelope mass before the hydrogen mixing.

The evolution during the central helium burning phase is exclusively determined by the size of helium core, and the enrichment of surface helium plays a part through the growth in the core mass due to hydrogen-shell burning. 
The mass loss through the RGB is also involved through
the reduction of envelope mass which influences the
hydrogen-burning rate as well as the effective temperature.  
    In the followings, we first discuss the former effects, and then, turn into the problem of mass loss.  

\subsubsection{The Luminosity and Lifetime on the HB}
Figure~10 shows the variations of the helium burning rate against the core mass interior to the hydrogen burning shell for the various mixing models. 
   Since the helium burning rate is a monotone increasing function of the helium core mass, the continuous mixing models start with larger helium burning rates than the no-mixing model.  
   For larger helium mixing rates, the helium burning rate is initially larger, and furthermore, increases during the core helium burning, to reach higher owing to the growth of core mass as well as to the reduction of helium abundance in the convective core.  
On the other hand, the intermittent mixing and the He-FDDM models begin with the helium burning rates, smaller than the no-mixing model and decreasing for greater helium enrichment, opposite to the continuous mixing models.  
   For all the modes of helium mixing, the helium core undergoes larger growth, and by the end of core helium burning phase, becomes more massive than for the no-mixing model, as a result of larger rates of hydrogen shell burning (see Figs.~2, 5, and 8).  
   Eventually, the models with greater helium enrichment end with larger helium core, and in particular, those starting with smaller initial core masses exhibit greater increment of core mass when compared among those with the same helium enrichment.   

The difference in the initial helium burning rate reflects on the lifetime of horizontal branch phase.  
   Figure~11 shows the time variations of mass in the hydrogen-depleted core through the horizontal branch for the various mixing models. 
   Among three modes of helium mixing, the continuous mixing models have the shortest lives on the horizontal branch or the shortest duration of central helium phase because of the initially largest helium burning rates, which decrease rapidly with the degree of helium enrichment.  
   On the other hand, the intermittent mixing and He-FDDM models may stay longer on the HB than the no-mixing model.  
The intermittent mixing first increases the HB lifetimes with the helium enrichment because of the decrease in the initial core mass, but it decreases for $\Delta Y_{\rm env} \ge 0.0793$ since the enhanced hydrogen shell burning rate accelerates the growth of core to raises the helium burning rate and hasten the central depletion of helium. 
The He-FDDM models have the longest lifetimes on the central helium burning because of the smallest core mass at the beginning, which monotonically increases with the helium enrichment.
  Because of longer HB lifetime and also of the concomitant CNO enhancement, these models experience the largest increment of core, much larger than the intermittent mixing models with the same helium enrichment. 

The resultant lifetimes on the HB are summarized in Figure~12 as a function of helium enrichment, and compare with the lifetimes on the RG brighter than the HB (here taken to be a fixed value of $L = 61.3 L_\odot$ for the ZAHB luminosity of no-mixing model).  
     For the continuous mixing, the HB lifetime is shorter than for the no-mixing model (by $\sim 12\%$ at the helium enrichment of $\Delta Y_{\rm env} = 0.0793$) and decreases with the degree of helium enrichment; 
     since the RGB lifetime increases with the helium enrichment, the ratio of the HB to RGB lifetimes decreases with the surface helium enrichment (by $\sim 20\%$ at $\Delta Y_{\rm env} = 0.0793$).  
     For the intermittent mixing, the variation in the HB lifetime is rather small, and hence, the ratio between the HB and RGB lifetimes may vary greatly with the mixing epoch; 
for an earlier occurrence of $\log (L/L_\odot) = 2.1$, the ratio of the HB to RGB lifetimes at $\Delta Y_{\rm env} = 0.0793$ is smaller by $\sim 33\%$ at $\Delta Y_{\rm env} = 0.0793$ than that for the no-mixing model, while the difference decreases to $\sim 10\%$ for a later occurrence of $\log (L/L_\odot) = 2.6$.  
    In comparison with the continuous mixing models, the intermittent mixing models give either smaller or larger ratios according to the difference in the mixing epoch.  
    Lastly, the He-FDDM models have the largest HB lifetimes, and hence, the largest ratios among all the mixing models and the no-mixing model.  
These ratios may have relevance to the observations of R-parameter, defined as the number ratio between the HB stars to the RGB stars brighter than the former \citep{Iben69};
for comparisons, we have to take into account the difference in the luminosity of horizontal branch since the duration of RGB sensitively depends on the assignment of the lowest luminosity, as will be discussed later.  
    
All the helium enriched models become brighter on the horizontal branch than the no-mixing model, including those with smaller cores, as seen from Figs.~1, 4, and 8, since the hydrogen shell-burning rate increases with the helium enrichment in the envelope and dominates the luminosity. 
     The dependence of luminosity on the horizontal branch is $\Delta \log L_{\rm HB} \simeq (2.1 \sim 2.2) \Delta Y_{\rm env}$ for the continuous mixing models, and slightly weaker $\Delta \log L_{\rm HB} \simeq (1.6 \sim 2.0) \Delta Y_{\rm env}$ for the intermittent mixing models because of smaller core.  
     The He-FDDM models have stronger dependence $\Delta \log L_{\rm HB} \simeq 2.3$ and $2.7 \Delta Y_{\rm env}$ because of the enhancement of CNO abundances with the helium enrichment. 
    For proper comparisons with the observations, however, we should take into account the change in the envelope mass due to the mass loss during the RGB evolution since the hydrogen burning rate depends on the envelope mass.  

\subsubsection{The effect of mass loss}
It has been known for long time that the observed horizontal branch stars require mass loss, and further, that variations in the mass loss rate may be counted for the second parameter necessary to explain the morphology of horizontal branch \citep{Cast69,Iben70,Rood73}.
   It is true, however, that a reliable theory of mass loss is not yet available to predict the reduction of envelope mass during the RGB phase \citep[e.g.,][]{Cate00}.  
Furthermore, in applying to the stars in the globular clusters, we have to take into account mass loss, or mass transfer, during the close encounters with other stars, as we have pointed out already, which may vary from star to star and from cluster to cluster according to their physical conditions. 
Here we here start by adopting the Reimers formula
\citep{Reim75} as a thumb rule to see an averaged effect of mass loss, i.e.,   
\begin{equation} 
\dot{M} = \eta \ 4 \times 10^{-13} \frac{L/L_\odot \cdot R/R_\odot}{M/M_\odot} (M_{\sun}/ \hbox{yr}) .
\label{reimers}
\end{equation} 
   We compute the evolution of no-mixing model, taking account of mass loss according to the above formula with $\eta = 1/3$ \citep{Reim77}, to find that the total mass reduces to $M = 0.6770 \msun$ by the onset of helium flash.  
   Apart from the reduction in the envelope mass, the mass loss little affects the evolution through the RGB, and in particular, leaves intact the core mass at the tip of RGB and the progress of helium core flash. 

Figure~13 illustrates the variations of total mass due to the mass loss against the core mass, computed from the luminosity and radius of the models without the mass loss; and the amount of the envelope mass, $\Delta M_{\rm ml}$, lost during the whole RGB evolution is listed in the last column of Table~1. 
The mass loss becomes important as the star climbs the RGB and expands since the Reimers formula (\ref{reimers}) implies the amount of mass loss in proportion to the surface radius if we take into account the fact that the duration of an evolutionary stage (while the core mass increases by a given amount) is in inverse proportion to the nuclear burning rate. 
The continuous mixing promotes the mass loss so that the integration of lost mass through the tip of RGB increases for larger helium enrichment, as discussed by \citet{Swei97a}, because of larger radii reached at the tip of RGB along with the elongation of RGB lifetime. 
The intermittent mixing, however, not necessarily shows the enhancement of mass loss in correlation with the degree of helium enrichment. 
   Instead, the intermittent mixing causes large variations in the amount of mass loss with the timing of mixing epoch.  
   As mixing epoch occurs later on the RGB at larger radii, the amount of envelope mass lost during the period necessary to produce the helium mixed into the envelope is larger. 
   Between our models with two different mixing epochs, the amount of envelope mass ejected during the mixing epoch increases from $0.013 \msun$ to $0.023 \msun$ for the helium enrichment of $\Delta Y_{\rm env} = 0.0793$. 
   Further an earlier occurrence of mixing epoch entails the ignition of helium core flash at smaller core mass, and hence, smaller amount of mass loss. 
   The both effects cooperate to make a difference of $\sim 0.02 \msun $ in the total envelope mass ejected during the RGB between these two models, which may causes significant influences on the horizontal branch evolution.  
    Among the models with the same epoch, on the contrary, the above two works opposite direction to each other with the variation in the degree of helium enrichment.  
    For larger helium enrichment, the production of helium spent for the surface enrichment takes longer time and increases the mass loss, while the resultant larger helium abundance in the envelope accelerates the evolution after the mixing epoch to ignite the helium core flash at smaller core masses with smaller radius and reduces the mass loss.  
    Accordingly, the mass loss is rather weakly dependent on the degree of helium enrichment, or even may decrease for larger helium enrichment.  
Here we evaluate the amount of mass loss by assuming that helium is produced as much as carried into the envelope during the mixing epoch at a constant core mass.  
   It is possible that shorter amount of helium can be produced, and hence, the core mass decreases during the mixing epoch if the inward mixing of hydrogen is very efficient. 
   Even in the latter case, the reduction of core mass remains smaller than $\vert \Delta M_1 \vert < \Delta Y_{\rm env} (M - M_1) / 0.96$ and will not affect so much our discussion.  
These characteristics, though derived by using the Reimers formulae, hold to all other formulae for the mass loss as long as they give the mass loss rate increasing more rapidly than the luminosity. 

In conclusion, the intermittent mixing and He-FDDM models can produce the variations in the envelope mass among the models with the same degrees of helium enrichment as large as those of the continuous mixing models with the different degrees of helium enrichment.  
In Fig.~9, we compare the variations in the total mass after the mass loss during the RGB and in the core mass at the onset of the HB as a function of the surface helium enrichment.
   The continuous mixing models enter into the horizontal branch with the envelope mass decreasing with the degree of helium enrichment mainly because of the increase in the total amount of mass loss during the RGB evolution.  
   On the contrary, the He-FDDM models show an opposite of large increase in the envelope mass with the mixing degree because of the largest decrease in the core mass.   
   The intermittent mixing brings about milder decrease in the envelope mass with the amounts between the two extremes of the continuous mixing and the He-FDDM, and covers over the significant range of envelope mass by changing the timing of mixing epoch for the same helium enrichment.   

The above results show that the continuous mixing models have the smallest initial envelope masses on the horizontal branch among the models of three mixing modes with the same helium enrichment.  But it may be the artifact that we consider only the mass loss, as described by Reimers-type formula.  
   In actuality, we should take into account an extra mass loss during the close encounter, which is postulated in the scenario for the hydrogen flash-assisted deep mixing and the He-FDDM, as stated in \S 2.  
Although we are concerned only with the relatively distant encounters, not so destructive as to strip the envelope off the red giants, it is shown that such tidal interactions as deposit a significant amount of angular momentum into the envelope of red giants are attended also with a loss of envelope mass up to an order of $0.01 M_\odot$, including the transfer onto the companion and/or the ejection from the system \citep[Yamada et al.\ in preparation, 2005]{Davi91}.  
    This additional losses of envelope may yield a spread in the envelope masses, and yet, because of the statistical nature of encounters, the mutual relationship between the intermittent mixing and He-FDDM models will be preserved with the averaged envelope masses decreasing with the later occurrence of mixing epoch for the intermittent mixing and the largest envelope masses for the He-FDDM model.   

In order to elucidate their relations to the mixing modes, we compute the evolution of the model stars with the mass loss taken into account by the Reimers formula with $\eta = 1/3$. 
   Although the choice of $\eta$ is rather arbitrary, these models may give an insight into the effects of mass loss on the horizontal branch morphology and of variations in the core masses, combined with the helium mixing.   
    Figure~14 shows the trajectories on the HR diagram after the models reach the horizontal branch for the different modes of helium mixing, and the duration of each part of their trajectories can be read from the time variations of effective temperature in Figure~15.  
   As helium is burnt in the convective core, the star first shrinks to raise the effective temperature owing to the increase in the central mean molecular weight, and then, expands to lower the effective temperature as the helium abundance in the convective core decreases significantly ($Y_{\rm c} \simeq 0.24$ for the no-mixing model) and the central temperature increases \citep{Sugi00}. 
As helium is exhausted in the center, the shell burning starts to expand the helium shell in the core and dwindle the hydrogen shell burning, contracting the star again to raise the effective temperature \citep{Haya62}, and then, the star expands as the mass in the helium-depleted core increases to evolve toward the asymptotic giant branch.  

The continuous mixing model starts the evolution of central helium burning from the bluest position than the no-mixing model and the other mixing models of the same $\Delta Y_{\rm env}$ because of the smallest envelope mass;  
   for greater helium enrichment, the initial envelope mass decreases monotonically (from $M_{\rm env} = 0.1724$ to $0.0166 \msun$ between $f = 0.1$ and 0.6), and the trajectory shifts more blue-ward, as already pointed out by the earlier works \citep{Swei97a,Swei98}. 
In addition, since the envelope mass may also affect the hydrogen burning rate in the shell surrounding the helium core, the continuous
models with the mass loss reduces the luminosity on the horizontal branch significantly as compared with the models computed earlier without mass loss in Fig.~1.  
Accordingly the mass loss offsets the effects of increasing luminosity with the helium enrichment.
For the case of $f \ge 0.3$ with the initial envelope mass $M_{\rm env} = 0.1281 \msun$, the luminosity tends to decrease despite the increase in the helium enrichment; 
   the models of $f \ge 0.4$ with $M_{\rm env} = 0.0963 \msun$ become even less luminous than the no-mixing model while the models without the mass loss are brighter by $\Delta \log L \gtrsim 0.45$.  
   Smaller hydrogen burning rate also entails smaller growth of helium core, and longer HB lifetime, as shown in Fig.~11.  
As compared with the models without the mass loss, the growth of core is much depressed and no longer monotone increase function of the helium enrichment;  
   the increment of core mass during the core helium burning is $\delta M_1 = 0.0246$, 0.0306, 0.0372, 0.0421, and $0.0366 \msun$ for $f = 0$, 0.1, 0.2, 0.3, and 0.4, respectively, while $\delta M_1 = 0.0352$, 0.0467, 0.0634, 0.0900, and $0.1330 \msun$ for the models without the mass loss.  

In contrast, the intermittent mixing models stays on redder side of the continuous models with the same helium enrichment.  
    Initially, the effective temperature little depends on the surface helium abundance because of massive envelope ($M_{\rm env} \gtrsim 0.19 \msun$), but later, the models of larger helium enrichment display a larger blue-ward shift as the core mass grows to reduce the envelope mass.  
The increment in the core mass is significantly larger than those of continuous mixing models with the same helium enrichment because of larger initial envelope mass, though much smaller as compared with the models without the mass loss ($\delta M_1 = 0.0494$ and $0.0420 \msun$ vs $\delta M_1 = 0.0685$ and $0.0643 \msun$ for the models of different mixing epochs with $\log L /\L_\odot = 2.1$ and 2.6 and with the same helium enrichment, $\Delta Y_{\rm env} = 0.0793$, computed under the assumption of the maximum cooling. See also Fig.~14).  
More remarkable is a variation in the trajectory due to the different mixing epoch, however.
   The model with the same helium enrichment reaches higher effective temperature as it has experienced the mixing epoch later on the RGB.

The He-FDDM models give the lowest effective temperatures among the helium mixing models because of the largest reduction in the core mass due to the helium mixing;  
   they are all located on redder side of the intermittent mixing models with the earlier occurrence of mixing epoch. 
For the lower helium enrichment of $\Delta Y_{\rm env} = 0.0354$, the model stays on the redder side whole through the core helium burning.  
   On the other hand, for the larger helium enrichment as $\Delta Y_{\rm env} = 0.1313$, the model, even though starting with still smaller effective temperature, makes a much larger blue-ward turn in the late phase when the envelope mass is reduced below $0.14 \msun$; 
the increment of core mass is $\delta M_1 = 0.1517 \msun$, and much larger than $\delta M_1 = 0.05372 \msun$ for the model of the smaller helium enrichment.  

It should be noted that the resultant envelope is subject to uncertainties in mass loss, as stated above. 
    Although we find the relatively low effective temperatures of $T_{\rm eff} \lesssim 10^4$ K for the intermittent mixing and He-FDDM models, high effective temperatures as observed can be readily realized if we adopt a slightly larger values for $\eta$ and/or if we take into account extra loss of envelope mass due to close encounters with environment stars in the dense stellar populations of globular clusters.  
   The additional loss of envelope mass may cause a further blue-ward shift of evolutionary trajectory on HR diagram, and also, a statistical spread of the stars along the HB in the latter case. 

In summary, the behavior on the horizontal branch stars can differ with the mixing models even with the same helium enrichment because of differences in the core mass and in the envelope mass.  
     As pointed out by \citet{Swei97a,Swei97b} and \citet{Lang95}, the helium mixing causes the blue-ward shift on the horizontal branch.  
     Both the continuous mixing and intermittent mixing models share this tendency.  
     In addition, the intermittent mixing gives rise to variations in the location on the horizontal branch according to the timing of mixing epoch, shifting stars bluer-ward as they have undergone the mixing epoch later on the RGB evolution.  
     Finally, the He-FDDM keeps the stars on redder side of the intermittent mixing models.  
     Accordingly, even within the limited range of the surface helium enrichment, it is possible to generate a wide spread in the color on horizontal branch from red to blue with the combination of these mixing modes. 
In particular, from the intermittent and the He-FDDM, we may expect a variation in the color of horizontal branch stars not a monotone function with the timing of close encounters that the giants come across.
   As a star undergoes such close encounter as leads the hydrogen flash-assisted deep mixing in later stage on the RGB, the descendant has to be situated on bluer side of horizontal branch, as predicted from the intermittent mixing model.  
If, on the other hand, a star has experienced the close encounter so late near the RGB tip to undergo the He-FDDM, it reaches the horizontal branch on the redder side than predicted for the intermittent mixing models even with the early occurrence of mixing epoch.  
   The angular momentum injected into the envelope of red giants from the orbital motion at these encounters may have relevance to the fast rotation of horizontal branch stars and the bimodal distribution of their rotation velocities, as observed from the globular clusters with the abundance anomalies, as discussed below. 

\section{Conclusions and discussion}
We have formulated and classified the possible mixing mechanism(s) to enrich the surface helium abundance in globular red giant stars into three modes, i.e., the continuous mixing, the intermittent mixing (or the hydrogen-flash assisted mixing), and the helium-flash driven deep mixing.  
     All modes postulate some extra mixing mechanism(s) other than the thermal convection and molecular diffusion; the continuous mixing requires the mixing mechanism that can transport processed material outward from the top of hydrogen burning shell, and the latter two demand the mixing mechanism(s) that can carry down hydrogen from the bottom of hydrogen-rich layer into the top of helium core.  
     We have computed the evolution of stars of mass $0.8 M_{\sun}$ and the metallicity $[{\rm Fe}/{\rm H}] = -1.5$ with an 0.4 dex enhancement of $\alpha$ elements through the horizontal branch with the surface helium enrichment due to the different mixing modes taken into account and explore how these different mixing mechanisms influence the evolution along the red giant branch and the horizontal branch morphology.  

    Our main results are summarized as follows;  
\hfill\break 
(1) \ The continuous and intermittent mixing modes have different, or even opposite, effects on the growth of core and also on the mass loss along the RGB. 
   The continuous mixing delays the ignition of helium core flash until a larger core, and hence, larger luminosity, are reached for greater degree of helium enrichment. 
   On the contrary, the intermittent mixing ends the red giant evolution with the core masses smaller for larger helium enrichment.  
    The intermittent mixing may occur at any stage on the RGB and the helium core flash tends to ignite at smaller core mass for the occurrence of mixing epoch in an earlier stage of RGB evolution.  
   
The both modes of mixing prolong the lifetime on RGB because of the production of helium used for the surface enrichment.  
   For the continuous mixing, the total amount of mass loss during the RGB, as evaluated from the Reimers formulae, increases for larger helium enrichment.  
On the other hand, for the intermittent mixing, it is rather dependent on the timing of mixing and only weakly on the helium enrichment; the injected mass decreases for an earlier occurrence of mixing, the tendency of which ensues from the competition between the mass loss during the period to produce the extra helium and the acceleration of core growth after the mixing epoch.  
\hfill\break
(2) \  The helium-flash driven deep mixing (He-FDDM) occurs at the tip of RGB during the helium core flash and has nothing to do with the RGB evolution. 
For this mixing, the reduction in the core mass is the largest among three modes of mixing since there is no extra production of helium used for the surface enrichment. 
\hfill\break
(3) \ When arriving at the horizontal branch, the continuous mixing model has the smallest envelope mass while the He-FDDM model has the largest envelope mass, and the intermittent mixing model falls between these two extremes, for a given degree of helium enrichment;  the envelope mass decreases with the helium enrichment for the continuous mixing because of the mass loss, while increases for the He-FDDM because of the decrease in the core mass.  
These results are based only on the mass loss given by Reimers-type formulae, but in actuality, we have to take into account additional mass loss due to close encounters that giants undergo with environment stars, in particular, for the intermittent mixing and He-FDDM models.  
The latter may reduce further the envelope mass, and in addition, give rise to a statistically spread distribution of horizontal branch. \hfill\break
(4) \ In general, the luminosity and the effective temperature of horizontal branch stars show tendency to increase with the helium enrichment.   
But the most critical thing in determining these properties is the envelope mass when the star reaches the horizontal branch, which may eliminate or even cancel out the effects of helium enrichment to increase the luminosity for such small envelope mass that gives higher effective temperatures than the red giant branch ($T > 10^4$ K). 
The envelope mass ensues from the competition between the growth of core and the mass loss, which differs among the mixing modes, and also from the loss of envelope mass during close encounter of red giants with nearby stars in the dense stellar environment, which should play an important part.

It has been long known that the distributions of stars on the horizontal branch vary greatly from cluster to cluster in particular for mildly metal-poor globular clusters as studied here.  
So far, attempts have been made to explain the spread and blue-ward shift of stars observed along the horizontal branch in terms of the differences in their ages and of stochastic variations in the amount of mass loss, and recently, helium enrichment as well as merging of planetary systems are proposed as their cause \citep[e.g., see][]{Cate05}. 
     The present results suggest that the difference in the mixing modes and also in the mixing epoch for the intermittent mixing can be additional factors, which may be associated with close encounters with other stars that giants may experience in the dense stellar environment peculiar to the globular clusters.  
The different timing of mixing epoch gives rise to variations both in the amount of mass loss along the RGB and in the core mass at the ignition of helium flash, resulting in variation of blue-ward shifts and spread of stars along the horizontal branch, indifferently of the degree of helium enrichment.  

The unique feature of the timing of mixing epoch as parameter lies in the fact that it can generate not monotonic variation in the location of horizontal branch stars as a function of the time when the giants have experienced close encounter along the RGB evolution.  
     As the stars come across the close encounter and undergo the mixing epoch later along the RGB, they tend to have smaller envelope, and hence, shift into higher surface temperatures on the HB.   
On the other hand, if the stars suffer from a close encounter very near to the tip of RGB, they will undergo the He-FDDM and will be situated at the redder side of the horizontal branch than the intermittent mixing models.  
    This may have the relevance to the distribution of rotation velocity observed along the horizontal branch of globular clusters. 
    Through these close encounters, the giants may once gain a significant amount of angular momentum to generate intense differential rotation between the envelope and the helium core, which can trigger the inward mixing of hydrogen and invoke the flash-assisted deep mixing.  
    After undergoing the mixing event, however, they discharge the deposited angular momentum effectively through mass loss during the ascent of RGB, and hence, those stars which experience the intermittent mixing in an early stage of RGB evolution will lose much of the angular momentum to be slow rotators when settling on the horizontal branch.  
    Those stars which undergo the close encounter and intermittent mixing later on the RGB also get rid of the amount of angular momentum with larger amount of envelope mass ejected through the mass loss, and finally settle on bluer side of horizontal branch as slow rotators. 
On the contrary, those stars which are subjected to close encounter very near to the tip of RGB and undergo the He-FDDM reach the redder side of HB than the intermittent mixing models, without losing the angular momentum acquired through the close encounter. 
     The latter may correspond to very fast spinning stars found on the redder side of HB, and we may expect the bimodal distribution of stellar rotation along the horizontal branch.  
     The globular clusters with abundance anomalies (M13, M15, M92) show the bimodal distribution of rotation velocity along the horizontal branch, fast rotation of $v \sin i \gtrsim 30 \hbox{ km s}^{-1}$ is found only on the redder side of $T_{\rm eff} = 15,000$ K, while the slow rotation of $v \sin i \lesssim 10 \hbox{ km s}^{-1}$ distributes continuously from the red to the blue \citep{Pete95,Cohe97,Behr00a,Reci02,Behr03a}.  
     The fastest rotation implies $\Omega / \Omega_{\rm K} \gtrsim 0.01$ at the tip of RGB if we assume the conservation of angular momentum, and SPH simulations of close encounters of giants with the main sequence stars have demonstrated that these orders of angular momentum can be transferred into red giants during a single encounter \citep[][Yamada, Okazaki, Fujimoto 2005 in preparation]{Davi91}. 
     Consequently, our scenario and the present results explain only to the source of angular momentum necessary for the observed fast rotation but also the bimodal distribution of globular cluster horizontal ranch stars.  
     This interpretation can be tested by exploring the surface abundances of fast rotating HB stars since their surface abundances have to be enriched in C and N as well as in helium such that $\Delta X_{\rm CN} \simeq 0.058 \Delta Y_{\rm env}$. 

The possibility that the stellar populations in the globular clusters are subject to modifications through the interactions with densely populated, environment stars, has been long considered and argued for some of the imprints of dynamical evolution of clusters, left on constituent stars \citep{Hill76,Djor91,Fusi93}.  
    It is also true, however, that the point has been raised that the stellar encounters, as evaluated from the observed physical parameters of globular clusters \citep[e.g.,][]{Hill76}, may not be numerous enough to explain the observed anomalies \citep[e.g.,][]{Djor91,Shim03}.  
    For example, recently reported is a correlation between the number of close binaries observed in X-ray and the stellar encounter rates in clusters \citep{Pool03}.  
    On the other hand, it is also pointed out that for the relative frequency of blue stragglers to horizontal branch stars or to the red giant, no significant dependence is found on the cluster parameters including the expected collision rates, apart from a mild dependence on the central density \citep{Piot04}.  
These facts may indicate that simple comparisons will not be warranted between the stellar properties and the stellar interactions inferred from the current physical conditions.  
It should be noted, however, that stars with the anomalies
have finite lifetimes, longer than the dynamical timescales
of clusters, though shorter than the secular evolution of
clusters, and hence, that the hysteresis of dynamical
evolution may play a role in the observed properties of clusters. 
    It is well known that the globular clusters experience the gravo-thermal oscillations, as first demonstrated by \citet{Sugi83} \citep{Bett84}, and repeat the collapse and the subsequent expansion in the relaxation timescales but with wide variations in their intervals and amplitudes \citep[see e.g.,][]{Maki96}.  
    During the collapsed phase, the timescales of stellar interactions decreases and stars may go through the close encounters in the central part of high stellar density.  
    It may be true that only a small fraction of cluster stars are involved in the collapsed core of high density, where the specific collision rates increase significantly. 
    But as already pointed out by \citet{Sugi96}, it is important to take into account the mass segregation as well as stellar evolutions in discussing the effects of dynamical evolution of clusters on individual stars.    
   Indeed, the mass segregation may increase the proportion of more massive stars in the central part and enhance the number of encounters during the core collapse involving red giants.  
   These close encounters may result in either tidal capture or flyby.   
   Once the tidally captured binaries are formed, the giants have to experience close approaches and/or exchange encounters with environment stars at much shorter timescales because of the increase in the cross sections due to the resonance and exchange interactions over the isolated stars \citep[see e.g.,][]{Hut92}. 
   The latter further enhances the stellar interactions not only to modify the stellar characteristics and evolution but also to change the spatial distribution through the recoil and ejection.  
  As a result, the appearance of stellar components in the clusters may differ with the degrees of core collapse and of mass segregation. 
   In addition, it critically depends on the time elapsed from the core collapse because owing to the finite lifetimes of RGB and HB stars ($\sim 0.1$ Gyr), anomalous RGB and HB stars formed during the core collapse disappear after finishing their lives.   
    Blue stragglers are also likely to form through the same kinds of encounters, and yet, may have different histories from them because of rather longer timescales for their formation and lives ($\gtrsim 1$ Gyr).  
   Accordingly, the differences in the elapsed time from the latest core-collapse phase may be counted as a candidate for the second parameter, which leads to cluster-to-cluster variations in the abundance anomalies, observed from red giants, in the morphology of the horizontal branch, and also in the population of blue stragglers.  

As for as the rotation velocity of horizontal branch is concerned, it is found that field blue horizontal branch stars also show bimodal distributions, similar to those observed from BHB stars in some clusters  \citep{Pete83b,Kinm00,Behr03b,Carn03}. 
   This forms a sharp contrast with the fact that the abundance anomalies are observed only from the cluster giants but not from the field giants. 
   Our scenario discussed above implies that the different source of angular momentum has to be assigned to fast rotating field BHB stars. 
   In fact, the angular momentum can be supplied through the spin-up of red giants in binaries due to the tidal locking of stellar rotation with the orbital motion, and the acceleration of rotation may also be possible through the merging of sub-stellar objects \citep[e.g., see ][]{Carn03}.  
   This surely works for the field stars in the Galactic Halo, but may not for stars in the globular clusters since the primordial binaries of such wide separation as can accommodate red giants are likely to have been disrupted through the interactions with dense environment stars in the globular clusters before the components evolve to giants.  
The synchronization of red giants proceeds in the timescales of stellar expansion or of evolutionary timescale of red giants, and hence, much more slowly than the angular momentum transfer due to the tidal encounters.  
   Accordingly, this may not generate strong differential rotations in the interior of red giant necessary to cause the inward mixing of hydrogen and produce the abundance anomalies as observed from cluster giants.  
   The binaries of separations less than the radius at the tip of RGB ($\sim 0.5$ au) undergo mass transfer when the primary stars evolve to RGB, to form blue stragglers.
   In actuality it is found that the frequency of blue stragglers relative to the horizontal branch stars are much higher among the field stars than in the globular clusters \citep{Pres00}. 
   These facts again indicate that the interactions with dense environment play critical roles in differentiate the evolution of stars in globular clusters from the filed halo stars.  
Proper understandings on the stellar evolution in globular clusters are yet to be established with influences of environments based on the dynamical evolution of clusters taking into consideration.

\subsection{Implications to Observations}
In this paper, we are devoted to reveal the basic properties of the different mixing modes and to formulate a framework to understand the abundance anomalies as observed from giants and the peculiar properties as observed from the horizontal branch stars at the same time.  
   We introduce the close encounters of giants with environment stars as a trigger of the mixing and propose the elapsed time from the last core collapsed phase during the gravo-thermal oscillations as the second parameter for the variations in the morphology of horizontal branch.   
Based on them, we argue the possibility that the observed
cluster-to-cluster variations can be interpreted in terms
of proper physical process in globular clusters.   

The formation of abundance anomalies in red giants in line with our scenario has been discussed \citep{Fuji99,Aika01,Aika04}. 
    As for the horizontal branch stars, the comparisons with the observations reduce to give a scheme representing their effective temperature and luminosity in terms of three parameters; the total mass, the surface helium enrichment, and the core mass, which we have demonstrated differ according to the mixing modes. 
   Although the detailed specification of the combinations of parameters should wait for future works, we here briefly review the implications of our scenario to observations by use of the present preliminary results in the following.  
   We address the problems of the lifetime on HB and on RGB and the age estimation of globular clusters. 
   In addition, we discuss about the luminosity functions of globular clusters and the second parameter problem.

Figure~16 shows the ratio of the lifetime of HB model to that of RGB model brighter than the luminosity level of the ZAHB as a function of degree of helium enrichment. 
   We take two kinds of reference luminosities of ZAHB, one fixed at the ZAHB luminosity ($L = 53.9 L_\odot$) taken from no mixing model with mass loss, and the other set at the ZAHB luminosity taken from each model.  
   These may approximate to the so-called R-ratio, defined as the number ratio of the HB stars to the RGB stars brighter than the luminosity level of the ZAHB \citep{Iben69}.  
   For both of the continuous and intermittent mixing models, the R-ratios decrease for constant reference luminosity owing to the decrease in the lifetime on HB and the increase of the lifetime on RGB.  
The reason for the increase of the RGB lifetime with
increasing $Y_{\rm env}$ is due to the time necessary
to produce the helium transported to the surface.
For the reference luminosities, taken from ZAHB of each model,
on the other hand, the R ratio increases to be little
dependent on the $Y_{\rm env}$ since the RGB lifetime
is mainly determined by the stage of lowest luminosity
and the lifetime ratio between the RGB and HB stars
results from the ratio of the consumption rates of
the nuclear fuels near these stages. 
   Lastly, the He-FDDM models have the largest R values among all the mixing models; their R-values are even larger than that of no-mixing model and increase with increasing helium abundance in the envelope even for the fixed reference ZAMS luminosity.   
   The R-value for the cluster as total may be given by the relevant combination of three modes or some of them.  
   In this figure, the cluster R values fall around 1.1 for a moderate amount of enrichment in the surface helium abundance, which is slightly smaller than those derived from observation of globular clusters for $\feoh \approx -1.5$ \citep[see for example, Fig.1 of][]{Cass03}.
   It should be noticed, however, that the inclusion of convective overshooting or semi-convection will lengthen the HB lifetimes and expected to yield larger R values.   

The helium mixing may affect the age estimation of globular clusters. 
    Since the mixing models have brighter HB luminosities than those without mixing, the $\Delta V$ method, based on the difference in the luminosities between the HB and Turn-off models, tends to give smaller age estimates when the helium mixing models are applied.  
   From the relationship between the HB luminosity and the envelope helium enrichment, discussed in \S 4.2.1, we estimate the maximal changes, $\Delta M_{\rm V}$, in the absolute magnitude due to the helium mixing, as 
\begin{equation}
\Delta M_{\rm V} = -2.5 \frac{\partial \log L\sufhb}{\partial Y\sufenv}
\Delta Y\sufenv \approx - 0.18 \sim -0.4
\qquad \hbox{for } \Delta Y\sufenv = 0.04 \sim 0.08.   
\end{equation}
     In deriving the most right-hand side member, we adopt $\Delta \log L\sufhb \approx 2.0 \Delta Y\sufenv$ in the models without mass loss. 
   If we take into account the reduction of HB luminosity with decreasing envelope mass, this imposes an upper bound to the variation in the HB luminosity, and hence, the scatter of HB luminosity due to the helium enrichment may fall practically in the range of current observational errors except for large enrichment exceeding this range.  
    
This difference in the HB luminosity may be transferred into an age estimate, $\tau$, of globular cluster, derived by the $\Delta V$ method through the luminosity dependence of $\tau (\propto M / L_{\rm ms} \propto L^{(1-\alpha)/\alpha}$ with $L \propto M^{\alpha}$), giving the variation, $\Delta \tau$, in the age estimation as 
\begin{equation}
\frac{\Delta \tau}{\tau} = \frac{1 - \alpha}{\alpha}
\frac{\partial \log L\sufhb}{\partial Y\sufenv}
\Delta Y\sufenv \frac{1}{\log e} \approx 0.13 \sim 0.26
\qquad \hbox{for } \Delta Y\sufenv = 0.04 \sim 0.08.  
\end{equation} 
    In the most right-hand side member, we assume $\alpha = 3.5$ according to the standard value for the zero-age main sequence stars.  
    If the helium abundances of HB stars in M13 do not exceed those in M3 by 0.04, as pointed out by \citet{Calo01}, the difference in the age estimates is at most $\sim 13$ \% between models with and without the helium mixing, rather small to discriminate it from the current observations.  
   Furthermore the reference level of ZAHB may depend on the degree and frequency of helium mixing, which introduces uncertainties in adopting the fiducial ZAHB to compare with observations and affects the precise estimation of ages in the actual observations.   

Thanks to the high resolution of telescopes, RGB luminosity function is available from recent observations of globular clusters \citep[see e.g.,][]{Cho05} to be compared with theoretical models.  
    The duration of the stay of models at certain luminosity depends on the growth rate of helium core mass. 
    For continuous mixing models, the growth rate of core becomes smaller than that of no-mixing model after the onset of mixing while it becomes larger in the late stage of RGB evolution, as discussed above.   
    Accordingly, this mode of mixing may lead to the enhancement in the luminosity function in the lower part of RGB branch.  
  The resultant luminosity function is dependent on the starting point of mixing and subject to the variations in the mixing rate.  
   Although we assume it to begin at a fixed luminosity, it is likely that the continuous mixing is also consequent upon the injection of angular momentum into the envelope of giant due to the close encounter, following the intermittent mixing.    
    The intermittent mixing may have a stronger tendency toward the enhanced luminosity function in the lower luminosity since the stagnation due to the production of extra helium during the mixing epoch and the acceleration of core growth after the mixing epoch.  
   In this case, however, we should take into account the frequency of encounter during the core collapse phase, which is larger for the giants of higher luminosity and of larger radius, and tends to shift the luminosity function toward higher luminosity.    
  The detail comparisons are awaited with these effects taken into account, which may constrain the modes of mixing and/or their occurrence.  

A comment is due here to the effects of star-star encounters in the globular cluster systems on the stellar distribution.  
    It may be thought that stellar encounter rates are to be peaked in the central core of globular clusters, and so their remnants. 
    In actuality, however, we should pay attention to their migrations after the star-star interactions. 
    Such close encounters as transfer significant amount of angular momentum into giants are likely to end in the formation of wide binaries and hence, because of large increase in the cross sections, the binaries formed are likely to undergo additional interactions with environment stars.  
    Eventually, through the resonant and/or exchange interactions, RGB stars in the binary can gain kinetic energy and be ejected from the core, and survive to be horizontal branch stars, systematically shifted toward bluer in H-R diagram.  
    It should be noted that because of the reduction in the mass during the encounter and due to the mass loss, the giants with abundance anomalies are no longer the most massive, as used to be except for the neutron stars, and are possibly ejected by the exchange interactions with more massive turn-off stars from the binaries, originally formed by tidal captures.  
    Since the crossing time scale of globular clusters ($\sim 10^7$ yr) is much shorter than their lifetimes ($\sim 10^8$ yr), such RGB stars as have experienced close encounters may escape from the core and be distributed in the whole cluster.  
    Therefore, strong radial concentrations toward the center may not be necessarily detected for the RGB stars with abundance anomalies and for the HB stars, as observed \citep{Bedi00}.
    As stated in \S1, this gives the mechanism of abundance anomalies, observed from dwarf and sub dwarf stars.  
    During the encounters, the giants suffer from an additional mass loss of envelope, which is accreted by the counterparts of encounters, mostly the main sequence stars.  

So far we have discussed the characteristics of each mode of mixing and the ways to discriminate them.  
    In actuality, they can be cooperative and/or complementary with each other.  
    The continuous mixing can readily produce the decline of carbon abundance during the ascent of RGB, but not the variations in the Mg and Al abundances to such large extents as observed.  
    On the other hand, the intermittent mixing can bring about the abundance anomalies of Mg and Al, including the reduction of \nuc{24}Mg. 
   These two processes may cooperate to produce the great depletions of oxygen since the intermittent mixing itself can produce the extents of abundance variations similar to that of \nuc{24}Mg, which is never observed to be depleted to not so large extent.  
   And yet, this does not necessarily imply that the continuous mixing can give rise to the surface helium enrichment;  
    in the mildly metal-poor stars, the oxygen burns in the top of hydrogen burning shell in which the amount of helium produced remains as small as $\Delta Y \simeq 3 X_{\rm C N} \simeq Z$ and it is possible to deplete oxygen with little helium enrichment \citep{Swei79}.  

As for the fast rotators, it is observed that they amount to approximately one third of the sample of horizontal branch stars in M13 \citep{Pete95}.  
    From our scenario, all these stars have to have experienced close encounters when they evolve near to the tip of red giant branch, and hence, have to occupy a fairly large fraction among the RGB stars.  
    It is yet to be properly investigated how efficiently the most massive component stars assemble into the central part and go through the encounters, and we should wait for the future simulations of stellar dynamics with the mass spectra and evolution of stars appropriately taken into account, and hence, can follow the mass segregation and the encounters of stars including the tidally formed binaries.  
    Here we point out that the collision rate with environment stars increases as the giants evolve to the tip, given the duration of high density due to the core collapse phase and also that the transfer angular momentum to the envelope is possible during such wide encounters as result in fly-by but not in the capture.  
    But for sufficient contributions from the He-FDDM, we may seek for the other source(s) of angular momentum for the fast HB rotators.  
    One of obvious candidates is the primordial binaries.  
    In this case, it is necessary to reveal the conditions that such wide binaries as allow the components to evolve to red giants can survive to present, which varies from cluster to cluster among those otherwise having the similar properties.

\acknowledgements
  We are very grateful to Dr. I. Iben for valuable comments and discussion and also for reviewing the manuscript. This work is in part supported by a Grant-in-Aid for Science Research from the Japanese Society for Promotion of Science (grant 15204010).  

\appendix
\section{Progress of helium enrichment for the continuous mixing}

In the continuous mixing models, the growth of helium core and the change in the chemical condition in the envelope due to the mixing are both related to the hydrogen burning rates, and hence, their rates are related as 
\begin{equation}
\frac{ d M_{1e}}{dt} = \frac{(1- f) L_H}{X_{\rm env} E_H} =  - \frac{1 - f }{ X_{\rm env} } \frac{1}{f} \frac{d X_{\rm env}}{ \ dt}(M - M_{1e}).    
\label{Xenv}
\end{equation} 
     In deriving the most right-hand side member, we apply eq.\ (\ref{eqY}).  
      For constant values of $f$, this yields an integration and the change in the surface abundance is given as a function of core mass; 
given as a function of the core mass $M_{1e}$;  
\begin{equation}
X_{\rm env} = 1 -Y_{\rm env} - Z_{\rm env}= X_0 [(M- M_{1e})/ (M - M_{1e, 0}) ]^{f/(1-f)}.  
\end{equation}
    Here we assume that the helium mixing starts at the stage of core mass of $M_{1e,0}$ with a uniform distribution of chemical composition of $X_{\rm env} = X_0$ and $Z_{\rm env}$ in the envelope. 
    The helium enrichment of numerical results differ slightly ($\lesssim 6\%$) from those evaluated from this equation because of a gradient of helium abundance in the envelope of the initial model.

\clearpage

\begin{table}
\begin{center}
\caption{Mixing models characteristics }
\begin{tabular}{lrrrrrl}
\tableline\tableline
model  &  $f$  &  $Y_{\rm env}$  &  $M_{1, \rm ZH} (M_\odot)$  & $ \log L_{\rm tip} (L_\odot)$ & $M_{2, i} 
(M_\odot)$ & $\Delta M_{\rm ml}$ \\
\tableline
No mixing        & 0.0 & 0.2491 & 0.4903 & 3.307 & 0.2194 & 0.1230 \\
\tableline
Continuous       & 0.1 & 0.2850 & 0.4913 & 3.344 & 0.2187 & 0.1363 \\
    Mixing       & 0.2 & 0.3284 & 0.4933 & 3.393 & 0.2277  & 0.1553 \\
                  & 0.3 & 0.3801 & 0.4945 & 3.444 & 0.2352  & 0.1774 \\
                  & 0.4 & 0.4438 & 0.4962 & 3.507 & 0.2493  & 0.2075 \\
                  & 0.6 & 0.6124 & 0.4960 & 3.642 & 0.2673  & 0.2874 \\
\tableline
Intermittent     & --- & 0.2850 & 0.4833 & 3.274 & 0.2042 & 0.1175 \\ 
Mixing           & --- & 0.3284 & 0.4765 & 3.293 & 0.1890 & 0.1153 \\ 
                 & --- & 0.3801 & 0.4706 & 3.303 & 0.1872 & 0.1158 \\
                 & 1.0 & 0.3284 & 0.4819 & 3.326 & 0.2112 & 0.1229 \\
                 & --- & 0.3284 & 0.4814 & 3.323 & 0.2118 & 0.1315  $\log(L/L_{\sun})$=2.6 \\
                 & 1.0 & 0.3284 & 0.4889 & 3.362 & 0.2417 & 0.1420  $\log(L/L_{\sun})$=2.6 \\
\tableline
Helium Flash     & 0.0 & 0.2845 & 0.4745 & 3.307 & 0.2194 & 0.1230 $X_{\rm mix}  = 0.001$ \\
Driven Mixing    & 0.0 & 0.3804 & 0.4211 & 3.307 & 0.2194 & 0.1230 $X_{\rm mix}  = 0.01$ \\
\tableline
\end{tabular}
\end{center}
\end{table}

\clearpage


\begin{figure}
\epsscale{.80}
\plotone{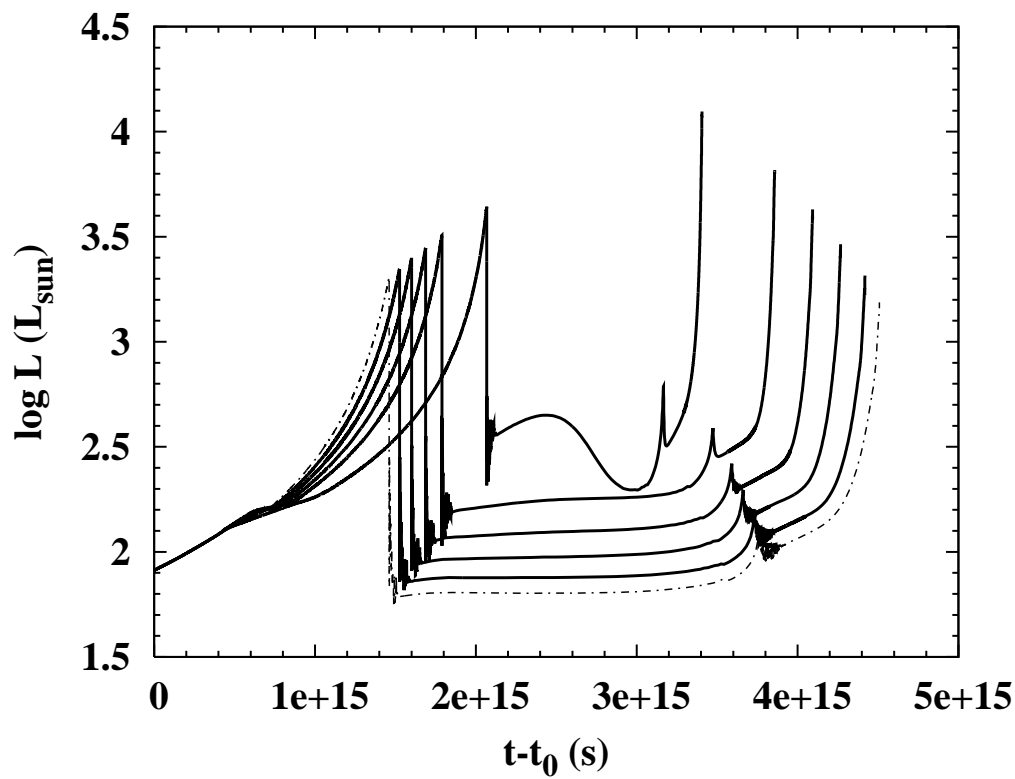}
\caption{Variations in the surface luminosity for the continuous mixing. 
   Solid lines denote the models with the mixing rate of $f = 0.1$, 0.2, 0.3, 0.4, and 0.6, from the bottom to the top, where the helium mixing is started at the stage of luminosity $\log L/L_\odot = 2.1$.  
     The dash-dotted line denotes the no-mixing model. 
     The origin of horizontal axis is taken to be the age of $t_0 = 4.86 \times 10^{17}$ s from the beginning of zero-age main sequence. 
\label{fig1}}
\end{figure}

\clearpage 

\begin{figure}
\plotone{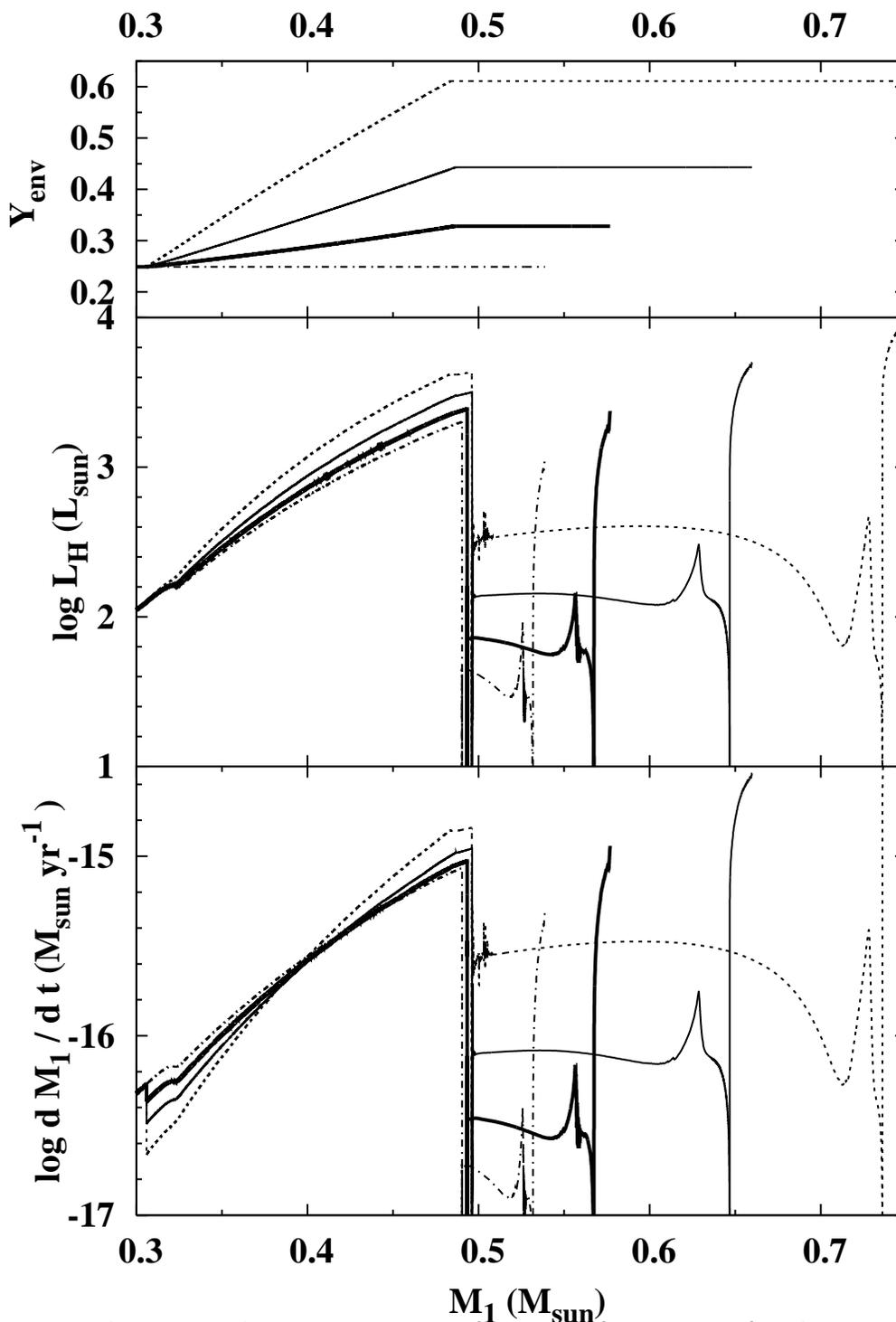}
\caption{Evolutionary characteristics as a function of core mass for the continuous mixing models with the mixing rate $f = 0.2$ (thick solid lines), 0.4(thin solid lines) and 0.6(broken lines) and for the no-mixing model (dash-dotted lines); 
   the surface helium abundance, the hydrogen burning rate, and the growth rate of core mass are shown, respectively, in the panels from the top to the bottom. 
\label{fig2}}
\end{figure}

\clearpage 

\begin{figure}
\plotone{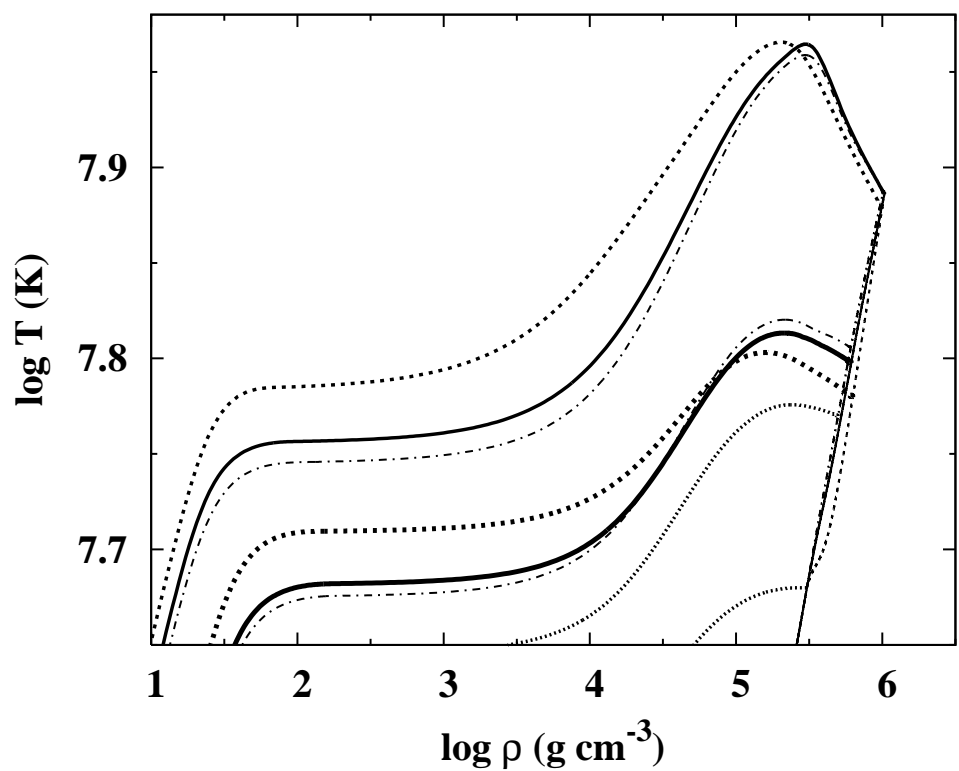}
\caption{The structure lines on the density-temperature diagram for the continuous mixing models with the mixing rate $f = 0.2$ (solid lines) and 0.6 (broken lines) and for the no-mixing model ($f = 0$; dash-dotted lines); thick and thin lines denote the structure lines at a stage of core mass $M_1 = 0.4 M_\odot$ (thick lines) and at the stage just before the ignition of major helium core flash (thin lines), respectively. 
     Two dotted lines denote those for the starting stages of helium mixing, $\log L /L_\odot = 2.1$ and 2.6.  
     Also plotted are the trajectories of the center of core up to the ignition of helium by the same lines as the structure lines. 
\label{fig3}}
\end{figure}

\clearpage 

\begin{figure}
\plotone{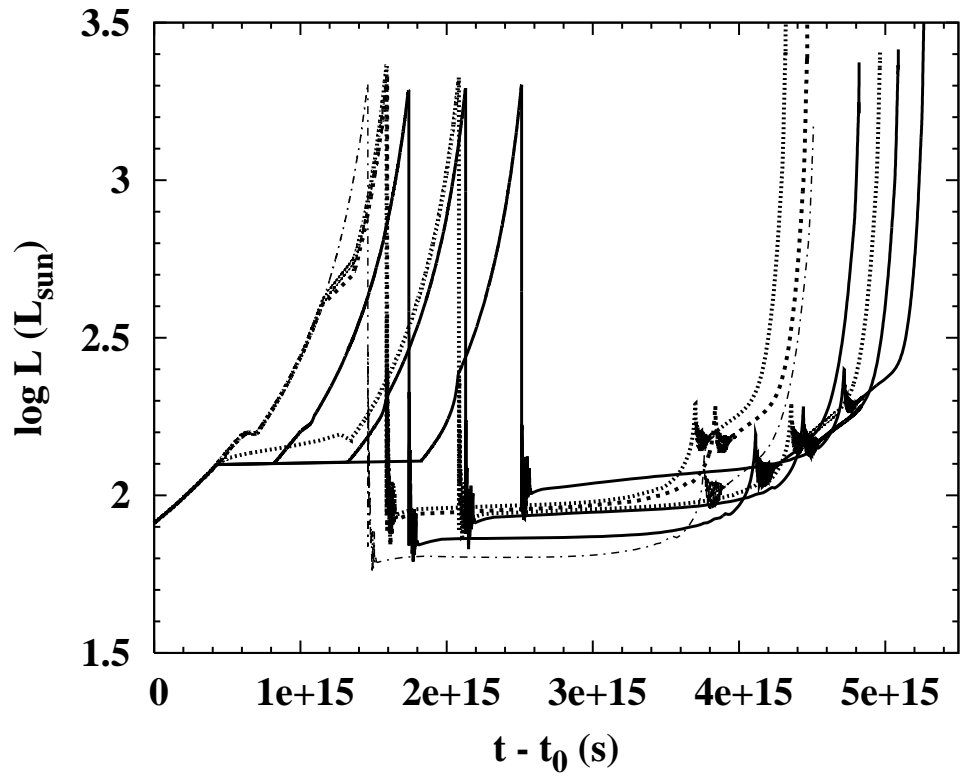}
\caption{
Time variations in the luminosity for the intermittent mixing models; solid and broken curves denote models with the mixing epoch at $\log L/L_\odot = 2.1$ and at $\log L/L_\odot = 2.6$, respectively.   
     Dotted curves denote those which take into account the time necessary for the helium production carried out to the surface, and also, the maximal cooling during the mixing epoch.   
     The no-mixing model is plotted by thin dash-dotted for comparison.  
\label{fig4}}
\end{figure}

\clearpage 

\begin{figure}
\plotone{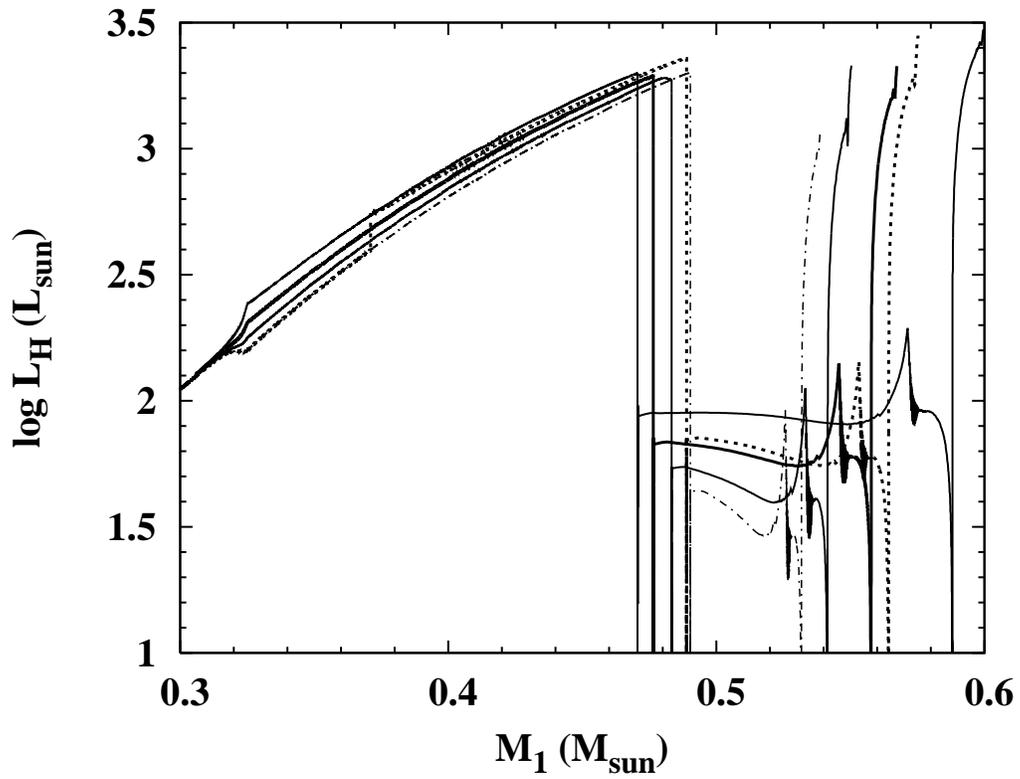}
\caption{
Variations in the hydrogen burning rate as a function of core mass for the intermittent mixing models; three models with the mixing epochs at $\log L/L_\odot = 2.1$ (solid lines) and a model with the mixing epochs at $\log L/L_\odot = 2.6$ (broken lines).   
     The no-mixing model is plotted by thin dash-dotted for comparison.  
\label{fig5}}
\end{figure}

\clearpage 

\begin{figure}
\plotone{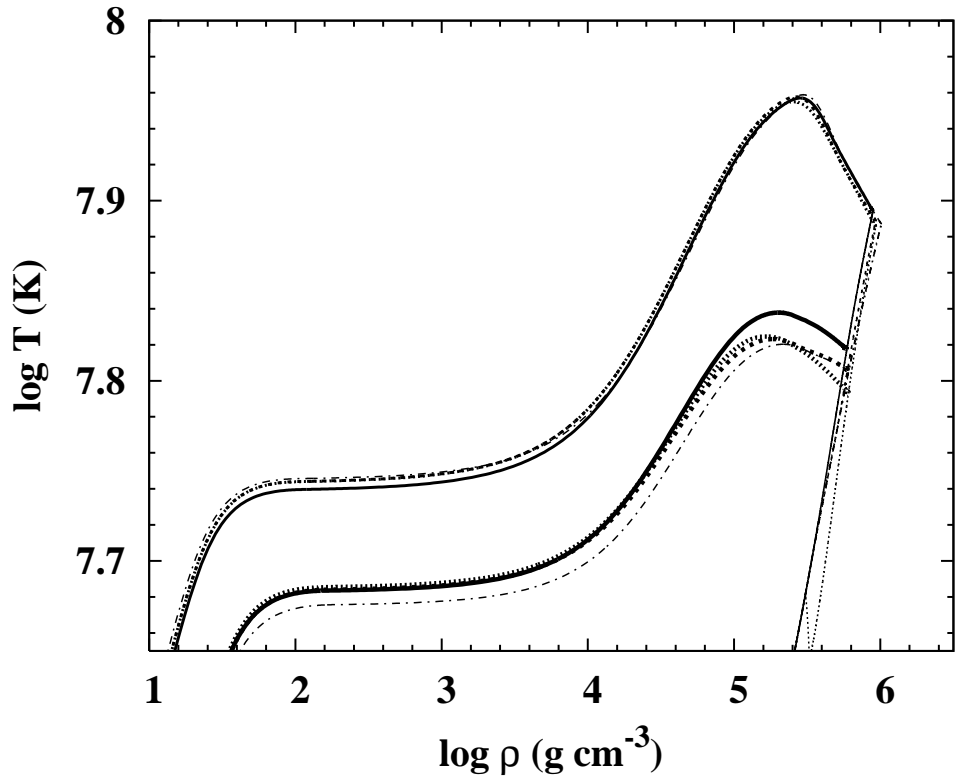}
\caption{
The same as Fig.~3 but for the intermittent mixing models
with the helium enrichment $Y_{\rm env} = 0.3284$:
solid and broken lines denote the models with the mixing
epoch $\log (L/L_\odot) = 2.1$ and with $\log (L/L_\odot) = 2.6$,
and dotted lines denotes the model with $\log (L/L_\odot) = 2.1$
and with the time to produce helium mixed into the envelope
and the maximal cooling during the mixing epoch taken into account.  
The model without helium mixing is plotted by dash-dotted lines for comparison.  
\label{fig6}}
\end{figure}

\clearpage 

\begin{figure}
\plotone{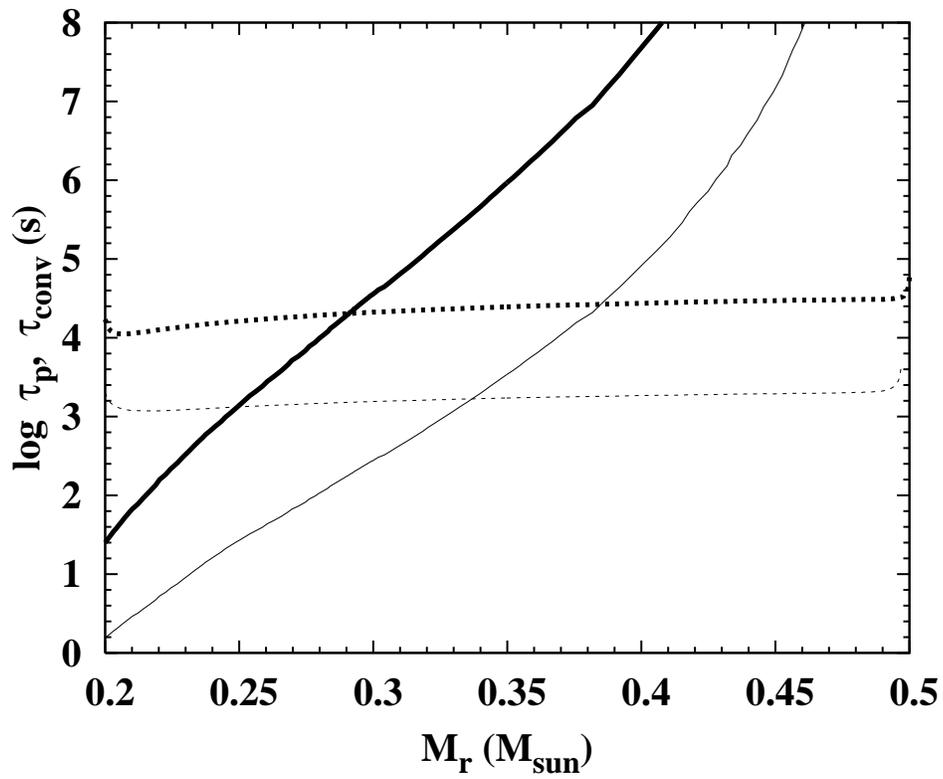}
\caption{
Lifetimes of proton, $\tau_{p}$, in the helium convection (solid lines)
and convective turnover timescales, $\tau_{\rm conv}$, (broken lines),
at the stage of the maximum extension of helium convection (thick lines)
and at a slightly earlier stage when the top of helium convection
comes within three pressure scale-heights to the hydrogen burning
shell (thin lines), plotted against the mass coordinate $M_r$. 
\label{fig7}}
\end{figure}

\clearpage 

\begin{figure}
\plotone{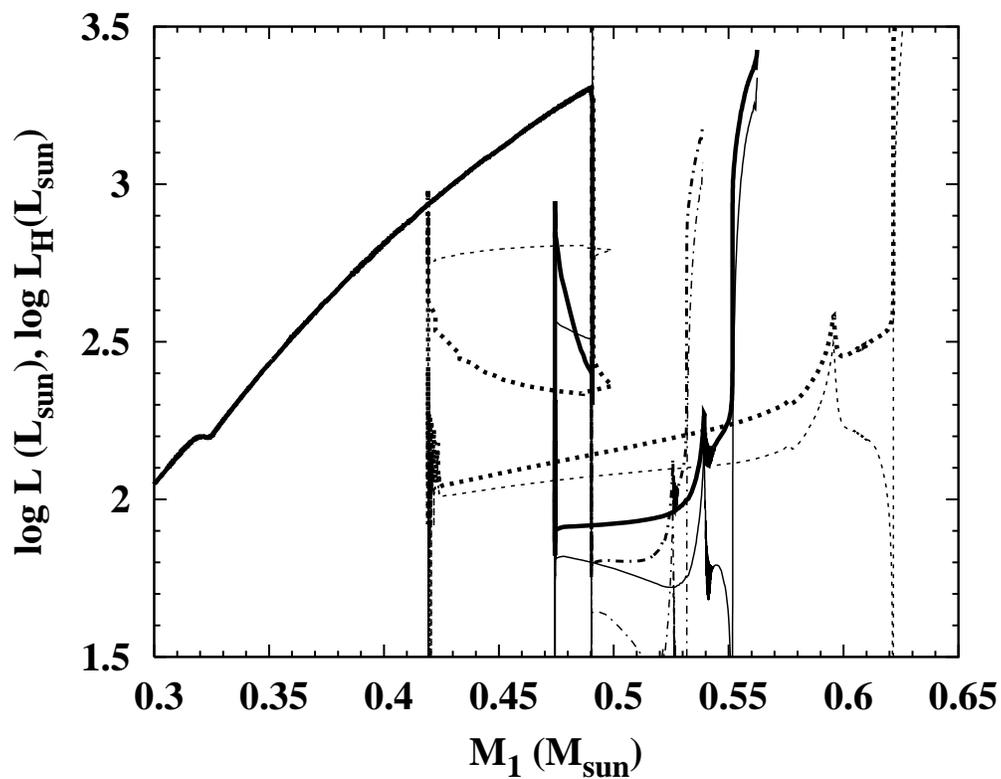}
\caption{
Variations in the luminosity (thick lines) and in the hydrogen-burning rate (thin lines) against the core mass for the helium flash-driven deep mixing models with the amounts of mixed hydrogen, $X_{\rm mix} = 0.001$ (solid lines) and $X_{\rm mix} = 0.01$ (broken lines). 
    Also shown by dash-dotted lines is the no-mixing model. 
\label{fig8}}
\end{figure}

\clearpage 

\begin{figure}
\plotone{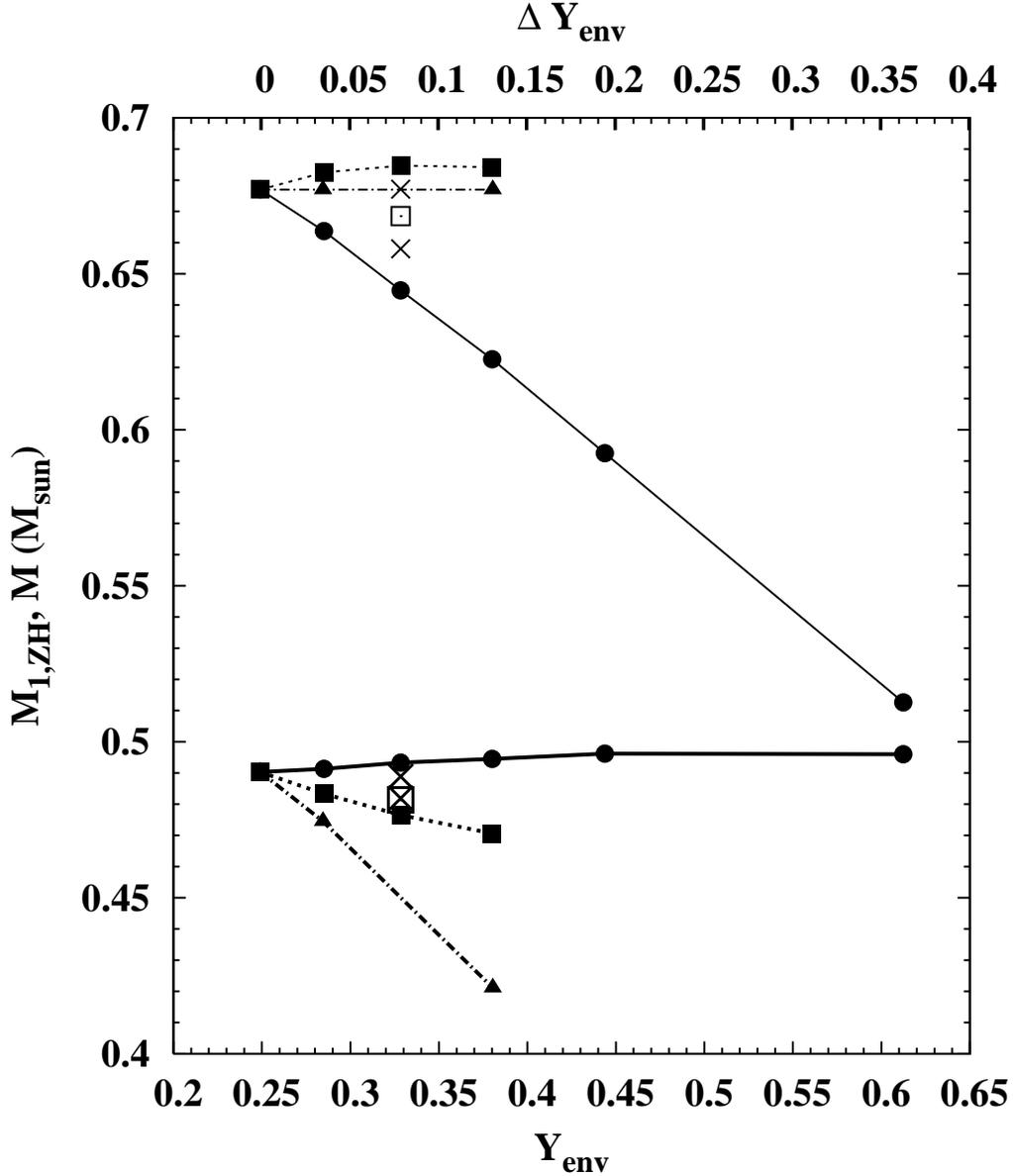}
\caption{
The core mass, $M_{1, \rm ZH}$ (bottom thick lines) and
the total mass, $M$ (top thin lines) at the beginning of horizontal
branch evolution, plotted against the helium enrichment, $Y_{\rm env}$,
for the three different modes of helium mixing, i.e.,
the continuous mixing (circle), the intermittent mixing with
the mixing epoch at $\log L/L_\odot = 2.1$ (filled square) and
at the mixing epoch at $\log L/L_\odot = 2.6$ (open square) and
with the maximal cooling during the mixing epoch taken into
account (crosses), and the helium-flash driven deep mixing (triangles).  
\label{fig9}}
\end{figure}

\clearpage 

\begin{figure}
\epsscale{.70}
\plotone{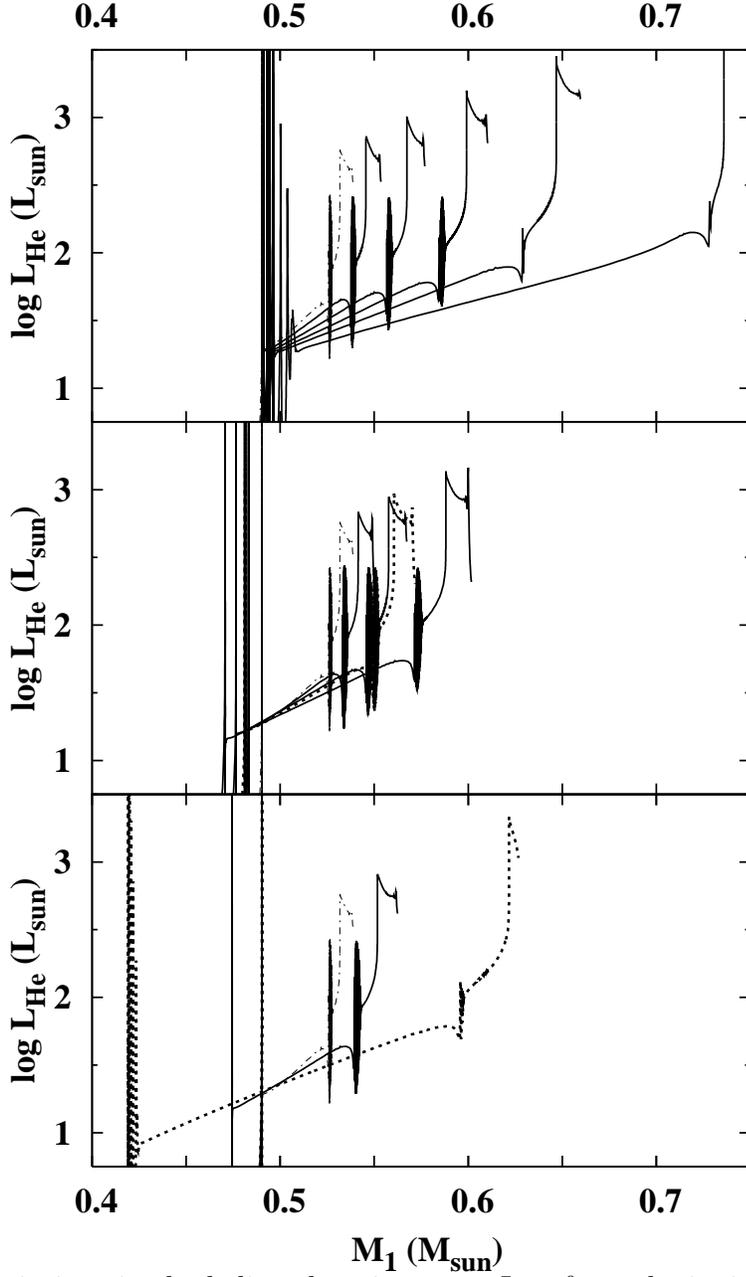}
\caption{
Variations in the helium burning rate, $L_{\rm He}$, from the ignition
of helium core flash through the central and shell helium burning and
to the beginning of thermal pulses on AGB, plotted against the mass
of hydrogen-depleted core, $M_1$, for the continuous mixing
models (top panel), for the intermittent mixing models (middle panel),
and for the helium flash-driven deep mixing (bottom panel).  
In the middle plane, solid and broken lines denote the intermittent
mixing models for the mixing epoch $\log L / L_\odot = 2.1$ and 2.6,
respectively. 
In the bottom panel, solid and broken lines indicate the models
with $X_{\rm mix} = 0.001$ and $X_{\rm mix} = 0.01$, respectively. 
Dash-dotted line in each panel denotes the no-mixing model. 
Vertical lines on the left end of each curve denote the helium
core flashes and the vertical segments densely overlapped in the
middle of each curve denote the oscillatory helium shell burning
that is shown to occur at the beginning of helium shell burning
around the helium-depleted core of mass $M_2 \lesssim 0.2 \msun$ \citep{Iben86}.
(Note that our computations do not take into account the
overshooting of convective core).
\label{fig10}}
\end{figure}
\clearpage 

\begin{figure}
\epsscale{.80}
\plotone{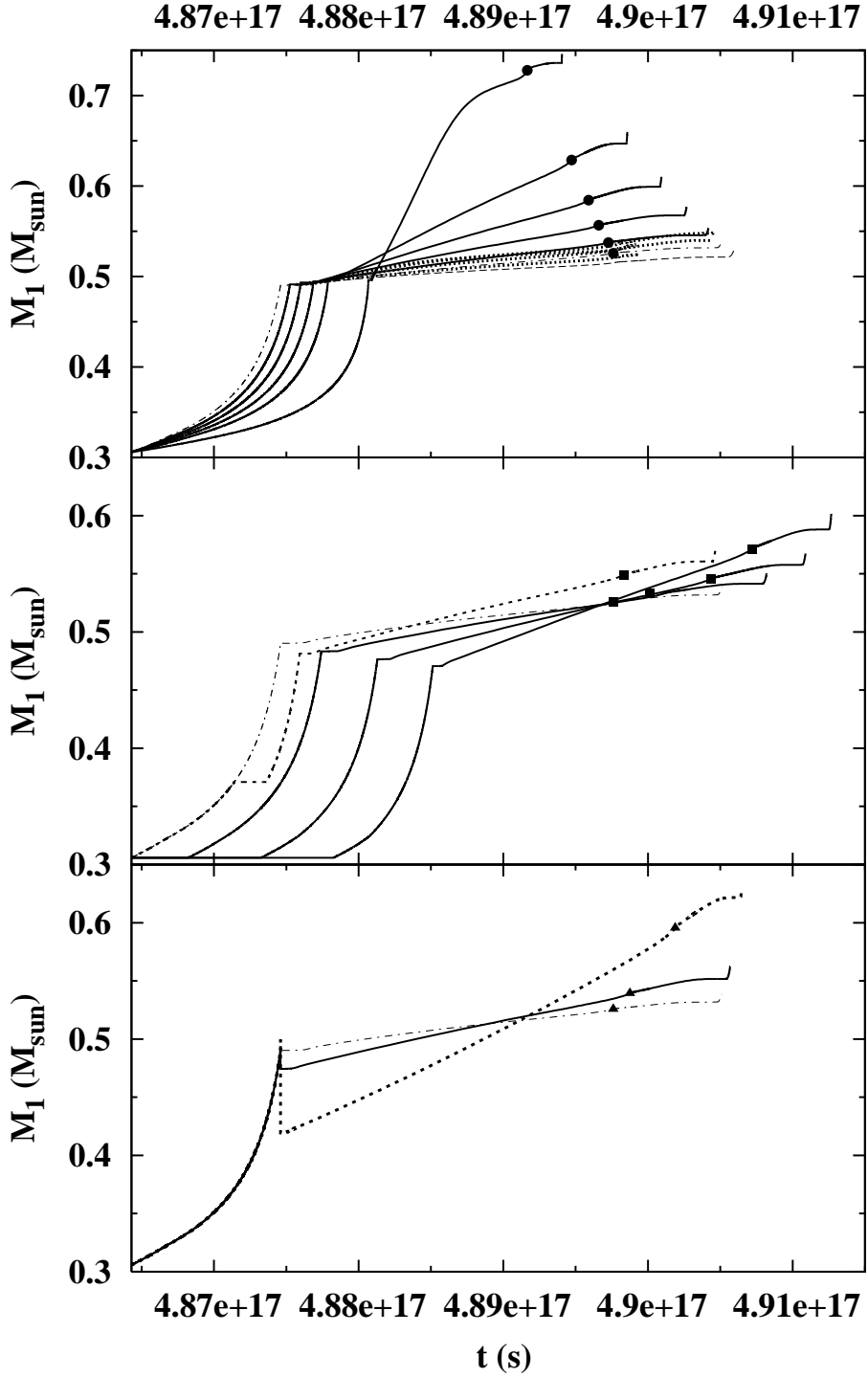}
\caption{
The time variations in the mass of hydrogen-depleted core for
the continuous mixing (top panel), the intermittent (middle panel)
and the helium-flash driven deep mixing (bottom panel).   
Designations of lines are the same as in Fig.~10, and symbols on
each curve denote the end of central helium burning.  
For comparison, the variations in the core mass when the mass
loss is taken into account are plotted for the continuous mixing
model by dotted lines and for the no-mixing model by thin dash-dotted line.  
\label{fig11}}
\end{figure}

\clearpage 

\begin{figure}
\plotone{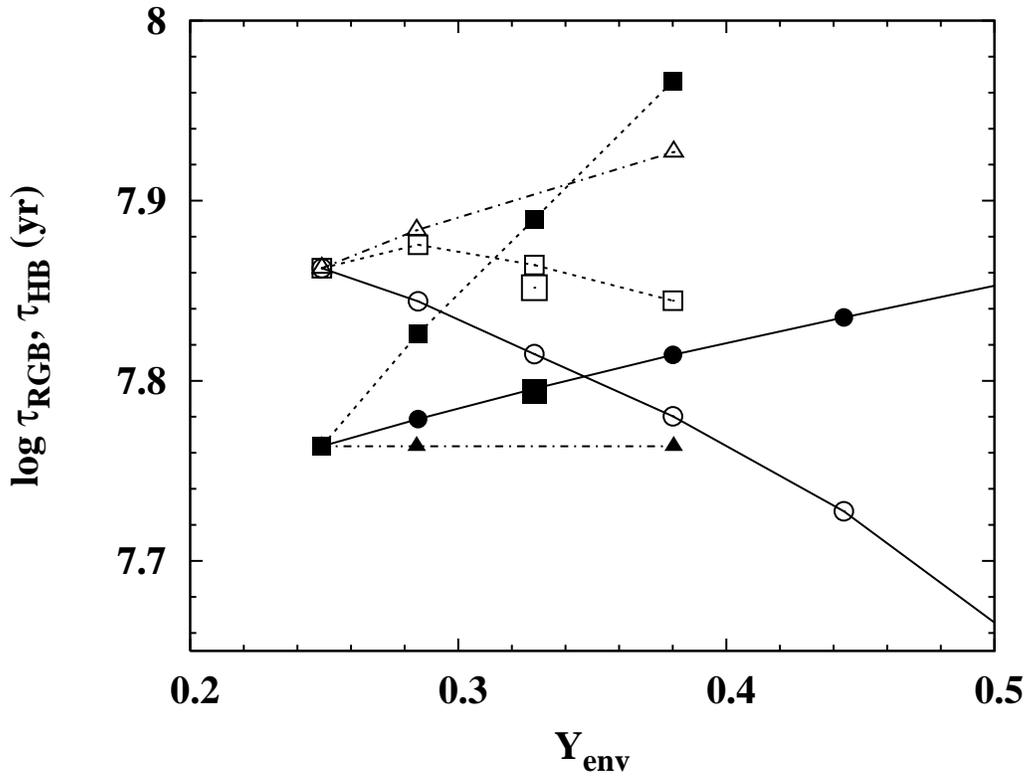}
\caption{
Lifetimes on the RGB (filled symbols) and on the horizontal branch
(open symbols) plotted against the degree of surface helium
enrichment for three modes of helium mixing, i.e., the
continuous mixing (circles connected with solid lines),
the intermittent mixing at the mixing epoch of $\log (L/L_{\sun}) = 2.1$
(squares with broken lines) and at the mixing epoch of
$\log (L/L_{\sun}) = 2.6$ (larger open and filled squares),
and the helium flash-driven deep mixing (triangles with dash-dotted lines). 
For the intermittent mixing models, the times necessary to
produce the amount of helium mixed into the surface are
added to the RGB lifetimes, estimated from the quiescent
hydrogen shell burning at constant core masses (see Fig.~11).  
\label{fig12}}
\end{figure}

\clearpage 

\begin{figure}
\plotone{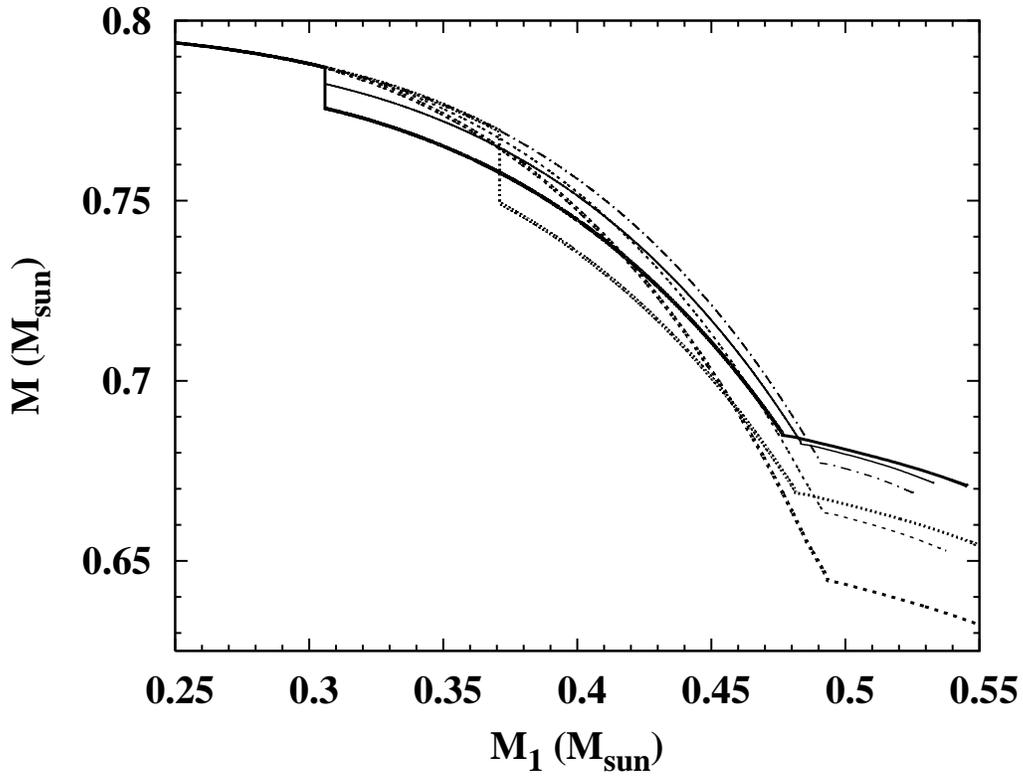}
\caption{
The decrease in the stellar mass due to the mass loss, computed from the Reimers formula with $\eta = 1/3$, plotted as a function of mass in the hydrogen depleted core for the no-mixing model (dash-dotted line), for the continuous mixing models (broken lines) and for the intermittent mixing models with the mixing epoch of $\log L / L_\odot = 2.1$ (solid lines) and with the mixing epoch of $\log L / L_\odot = 2.6$ (dotted line), respectively. 
    For the continuous mixing and the intermittent mixing with the earlier mixing epoch, two models are shown with different surface helium enrichment $\Delta Y_{\rm env} = 0.0359$ (thin lines), and 0.0793 (thick lines).   
\label{fig13}}
\end{figure}

\clearpage 

\begin{figure}
\plotone{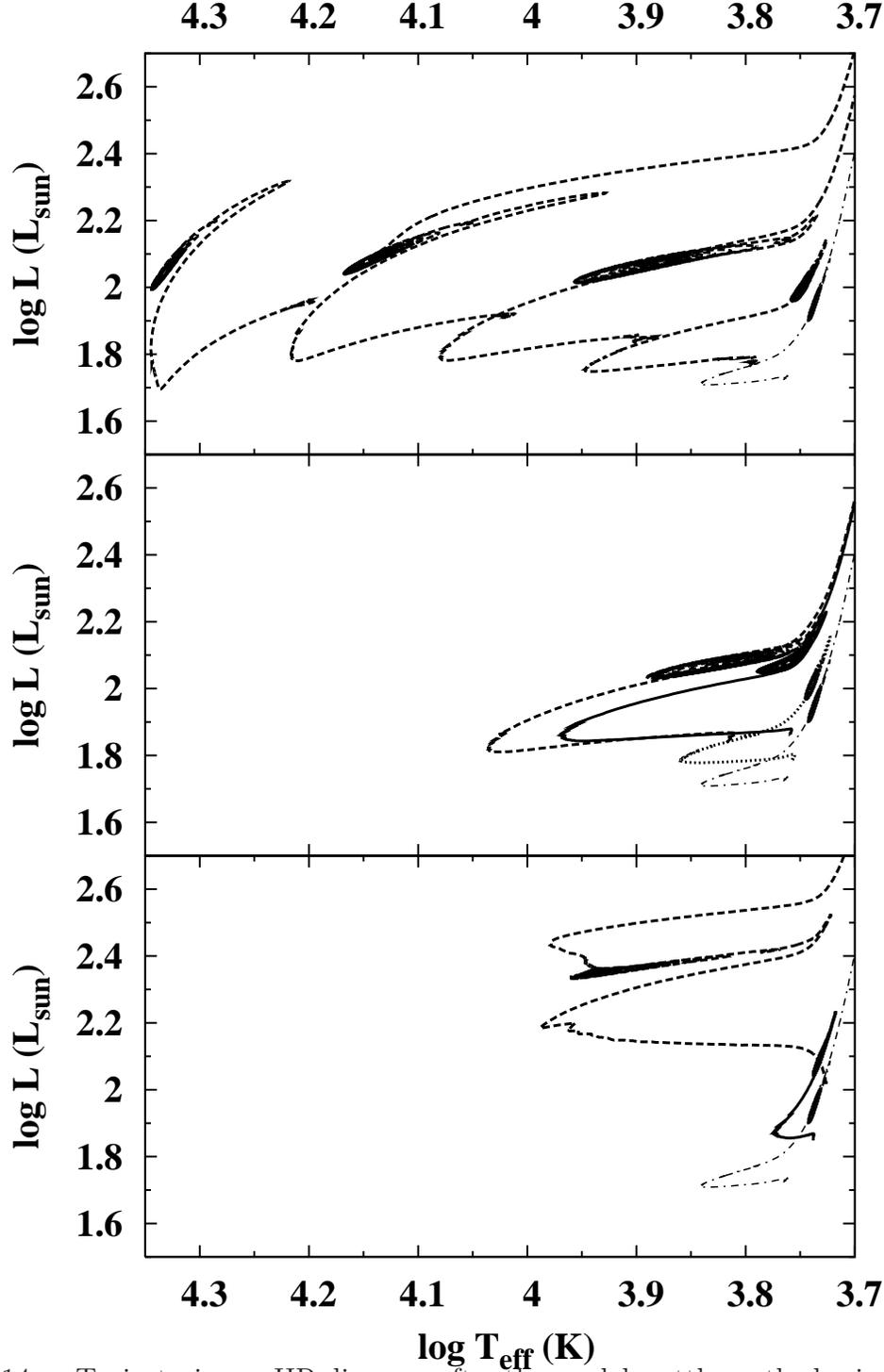}
\caption{
Trajectories on HR diagram after the models settle on the horizontal branch when the mass loss is taken into account according to the Reimers formula with $\eta = 1/3$.  
   Top panel for the continuous mixing models with $f = 0.1$, 0.2, 0,3, 0.4 from right to left:  
   middle panel for the intermittent mixing models computed with $f=1.0$ and $\Delta Y_{\rm env}=0.0793$ at mixing epochs $\log L/L_\odot = 2.1$ and 2.6 (solid and broken lines) and that of $\Delta Y_{\rm env} = 0.0359$ (dotted line):  
   and bottom panel for the helium-flash driven mixing models with $X_{\rm mix} = 0.001$ (solid) and 0.01 (broken line), respectively.    
     Dash-dotted line on each panel denotes the no-mixing model.   
\label{fig14}}
\end{figure}
\clearpage 

\begin{figure}
\plotone{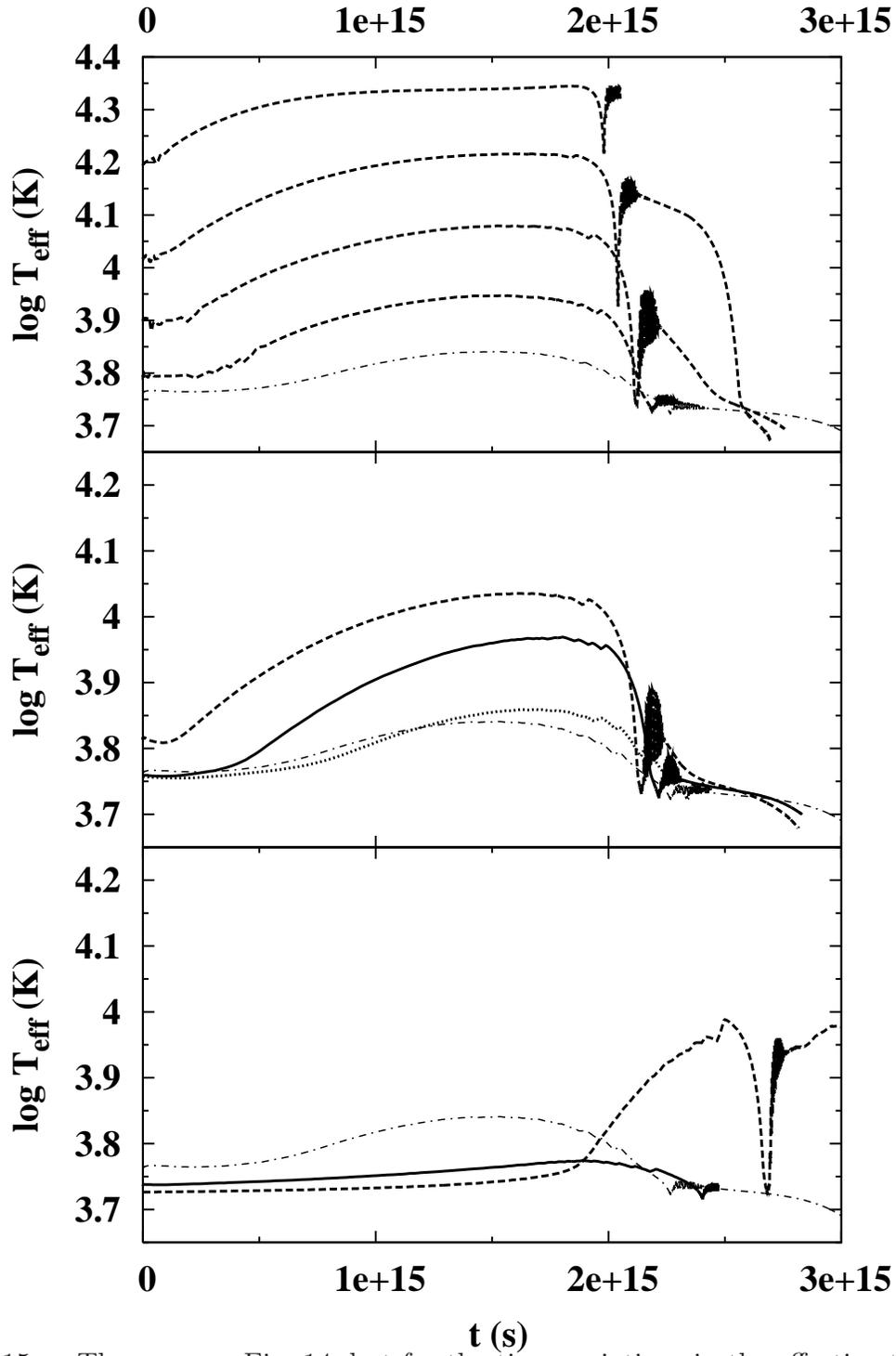}
\caption{
     The same as Fig.~14, but for the time variations in the effective temperature.   
\label{fig15}}
\end{figure}

\clearpage 

\begin{figure}
\epsscale{.70}
\plotone{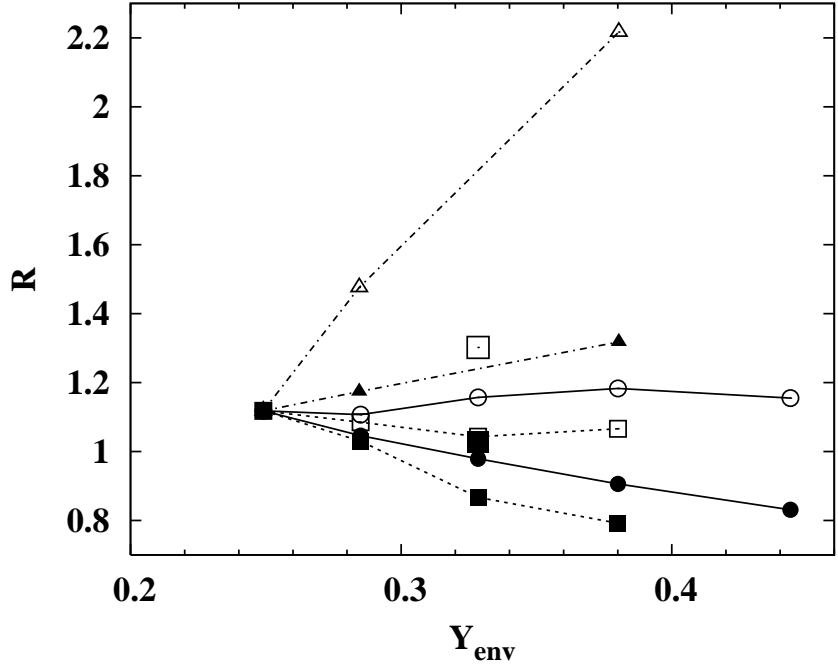}
\caption{
Ratio, $R$, of lifetimes on the horizontal branch to those on the red giant branch brighter than the reference luminosity, plotted against the degree of surface helium enrichment for the continuous mixing (circles connected with solid lines), the intermittent mixing at the mixing epoch of $\log (L/L_{\sun}) = 2.1$ (squares with broken lines), and the helium flash-driven deep mixing (triangles with dash-dotted lines).  
    Open and filled symbols denote the values when the reference luminosity is set at the luminosities of zero-age horizontal branch of the no-mixing model (i.e., $L = 53.9 L_{\sun}$) and at those of individual models, respectively;  
    large open and filled squares denote those for the intermittent mixing models at the mixing epoch of $\log (L/L_{\sun}) = 2.6$.  
    The lifetimes of horizontal branch are taken from the models with mass loss, differently from those in Fig.~12.
\label{fig16}}
\end{figure}

\end{document}